\documentclass[12pt]{iopart}
\usepackage{amssymb}
\usepackage{epsfig,color}
\def\eps{\varepsilon}
\begin{document}
\title{Faraday waves in BEC with engineering three-body interactions}
\author{F. Kh. Abdullaev$^{1,2}$, A. Gammal$^{3}$ and Lauro 
Tomio$^{2,4}$\footnote{{Corresponding author: tomio@ift.unesp.br}
}}
\address{ 
$^1$ Department of Physics, Faculty of Sciences, IIUM, Bandar Indera Mahkota, Jln. 
Sultan Ahmad Shah, 25200,  Kuantan, Malaysia.\\
$^2$ CCNH, Universidade Federal do ABC, 09210-170, Santo Andr\'e, Brazil.\\
$^3$ Instituto de F\'\i sica, Universidade de S\~ao Paulo, 05508-090, S\~ao Paulo, Brazil.\\
$^4$ Instituto de F\'\i sica Te\'orica, UNESP-Universidade Estadual Paulista, \\
01140-070, S\~ao Paulo, Brazil.
}
\date{\today}
\begin{abstract}
We consider Bose-Einstein condensates with  two- and three-body interactions periodically 
varying in time. Two models of time-dependent three-body interactions, with quadratic and quartic 
dependence on the two-body atomic scattering length $a_s$, are studied. It is shown that  
parametric instabilities in the condensate leads to the generation of Faraday waves (FW), with 
wavelengths depending on the background scattering length, as well as on the frequency and 
amplitude of the modulations of $a_s$. In an experimental perspective, this opens 
a new possibility to tune the period of Faraday patterns by varying not only the frequency of 
modulations and background scattering length, but also through the amplitude of the modulations.
The latter effect can be used to estimate the parameters of three-body interactions from the FW 
experimental results.  Theoretical predictions are confirmed by numerical simulations of the 
corresponding extended Gross-Pitaevskii equation.
\end{abstract}
\maketitle

\section{Introduction}

The role of three body-interactions in Bose-Einstein condensates (BEC) has attracted great deal 
of attention~\cite{Petrov,3bodyrev,Bulgac,Braaten,Kohler}.  Typically, in such kind of dilute system as BEC, 
three body effects are quite small in comparison with two body effects. {One of the relevant role that 
a three-body interaction can play was shown in Ref.~\cite{Gammal2000}, in the particular case 
of attractive two-body interactions, where a critical maximum number of atoms exist for stability. 
The addition of a repulsive three-body potential, even for a very small strength of the three-body 
interaction, can extend considerably the region of stability. It was also shown in 
\cite{Abdullaev-PRA2001} that, if the atom density is considerably high, the three-body interaction 
can start to play an important role. More recently, a possible interesting scheme for obtaining a 
condensate with  almost pure three-body effects has been suggested in \cite{Tiesinga}.} 
The idea consists in implement periodical variations in time of the $s-$wave atomic scattering length 
$a_s$ near zero, such that we have a varying two-body interaction. 
It can be achieved, for example, by using Feshbach resonance technics, varying the external 
magnetic field near the resonance~\cite{Inouye,Chin}. Therefore, by considering this procedure,
two-body effects can be averaged to zero, enhancing the effective three-body interaction, which
is proportional to an even power of the two-body interaction. Note that, an analogue of this 
scheme has been considered before, within an investigation of the role of three-body 
interactions to arrest collapse in BEC~\cite{AG}. 
At the same time, the periodic modulations in time of two-  and three-body interactions can 
lead to parametric instabilities of the ground state, resulting in the generation of Faraday 
waves~\cite{Faraday}. 
The Faraday waves (FW) are pattern in BEC which are periodic in space, with the period 
determined by the periodic modulation of trap 
parameters and strengths of the two- and three body interactions. The FW in BEC with two-body 
interactions has been investigated theoretically in \cite{LVS}, as well as created in cigar-shaped 
experimental setups~\cite{Engels}. 

Much attention has been devoted recently to FW in several investigations. They can be divided into 
two groups. First, by dealing with time variation of transverse trap parameters, leading to effective 
time-dependent nonlinearity in the reduced low-dimensional Gross-Pitaevskii (GP) equation.
As examples with oscillating transverse frequency of the trap, we can mention the following recent studies:
low and high density one-component BEC~\cite{LVS,Nicolin2,Nicolin3,Nicolin1,Nicolin};
two-component BEC in trap with modulated transverse confinement~\cite{Bhatt,Nicolin2}; 
Fermi superfluids in BEC with two- and three-body interactions~\cite{Tang}; and 
Fermi superfluids at zero temperature~\cite{Capuzzi}, where the FW were considered as 
a relevant tool to study BCS-BEC crossover. 
A second group of investigations have used the modulation in time of the strength of the interaction. 
As an example, we have two component BEC with time-dependent  inter- and intra-species interactions, 
with single and coupled  BEC's with time-dependent dipolar interactions, where FW are considered as 
excellent tool to study nonlocal effects in polar gases~\cite{Santos,Lakomy}. Another example is given 
in \cite{AOS}, considering studies of superfluid Bose-Fermi mixtures with modulated two-body 
scattering length for the bose-fermion system. Belong also to this group the works with analysis of 
parametric instabilities  in an array of BEC with varying atomic scattering length, based  on  a discrete 
nonlinear Schr\"odinger equation~\cite{Rapti}. One should also observe that 
analogue of FW patterns can also be found in optical fibre systems~\cite{ADCS,ADG,Fabio}.

By taking into consideration effects due to three-body interactions in BEC, an important point that one 
should consider is that three-body effects are defined by the value of the two-body interaction (atomic 
scattering length). Therefore, by varying in time the scattering length, the three-body interaction will 
also be affected, with the functional form being defined by the corresponding physical model.
By taking into account this dependence, one can expect new peculiarities in the FW generation, 
in BEC with two- and three-body interactions. With this motivation,  we are concerned to the 
present paper with an investigation of FW generation in BEC by considering two possible   
regimes leading to the modulation of the three-body parameter. First, motivated by a model presented
in Ref.~\cite{Tiesinga}, we analyse the case when the strength of the three-body interaction is 
proportional to the square of the two-body scattering length. 
In such model, the corresponding Gross-Pitaevskii type of equation has a  term that 
mimics three-body interactions,  which appears  at the description of high-density BEC in 
cigar-type traps~\cite{Salasnich,Khaykovich}.  
A second possibility for the modulation of the three-body parameter can arise by considering 
the case of large two-body scattering lengths, near the Efimov regime~\cite{Efimov},
where the number of three-body states (resonant or bound) increase as the energy of the
two-body system goes to zero. In such a case, the strength of the three-body interaction is 
predicted to be proportional to the fourth power of the atomic scattering length~\cite{Bulgac, Braaten}.  

In both the cases we have analysed, we observe that the FW parameters depend additionally 
on the amplitude of the time modulations of the atomic scattering length, not just on the corresponding
frequency and background two-body scattering length, such that one can experimentally tune the 
wavelength of FW. In this way, from the amplitude of the modulations necessary to obtain 
experimentally the FW patterns one can also  estimate the two- and three-body interaction parameter.

\section{The model}
\label{sec:mod}

Let us consider a quasi-one-dimensional Bose-Einstein condensate with atoms of mass $m$, with two- 
and three-body interactions varying in time. The system is described by a one-dimensional (1D) 
time-dependent Gross-Pitaevskii equation (GPE), with cubic and quintic terms parametrised, respectively, 
by the functions $\Gamma(t)$ and $G(t)$. 
By also considering a possible time-independent external interaction $V_{ext}$, with the wave-function
$\psi\equiv\psi(x,t)$ normalized to the number of atoms $N$, the equation is given by
\begin{equation}\label{eq01}
{\rm i}\hbar\frac{\partial\psi}{\partial t} =
-\frac{\hbar^2}{2m}\frac{\partial^2\psi}{\partial{x^2}} + V_{ext}(x)\psi
- \Gamma(t)|\psi|^2\psi - G(t)|\psi|^4\psi,
\end{equation}
where $\Gamma(t)$ is related linearly with the two-body $s-$wave atomic scattering length $a_s(t)$,
which can be varied in time by considering Feshbach resonance techniques\cite{Inouye}. The possible 
ways that the three-body strength $G(t)$ can be varied in time will depend on specific atomic characteristics, 
which are also related to the kind of two-body interaction, as well as induced by some external interactions 
acting on the condensate.

Several examples can be considered, following Eq.~(\ref{eq01}), which can be rewritten with dimensionless
quantities~\cite{AS05}, by changing the space-time variables such that $ t\to t/\omega_{\perp}$ and 
$x \to {x}{l_{\perp}}$, where we have a length scale $l_\perp$ and a transverse frequency $\omega_\perp$ 
related by $l_{\perp} \equiv \sqrt{{\hbar}/{(2m\omega_{\perp})}}$.
Therefore, in the new dimensionless quantities, with $u\equiv u(x,t) = \sqrt{l_\perp} \psi$ and
$V_{ext}=\hbar\omega_\perp {\cal V}_{ext}$, we have 
\begin{equation}\label{eq02}
{\rm i}\frac{\partial u}{\partial t} + \frac{\partial^2 u}{\partial{x^2}} 
- {\cal V}_{ext}(x) u + \gamma(t)|u|^2 u + g(t)|u|^4 u = 0,
\end{equation}
where the dimensionless time-dependent two- and three-body parameters are, respectively, 
given by 
\begin{equation}
\gamma(t)\equiv \sqrt{\frac{2m}{\hbar^2}} \frac{\Gamma(t)}{\sqrt{\hbar\omega_\perp}}
\;\;\;{\rm and}\;\;\;
g(t)\equiv \frac{2m}{\hbar^2} G(t)
\label{eq03}. \end{equation} 
In the following expressions, we consider that no external potential is applied to the 
system (${\cal V}_{ext} = 0$), such that the natural scale is the $s-$wave two-body scattering
length $a_s$ at $t=0$, which will define $\omega_\perp$ and the corresponding length
$l_\perp = 2a_s$.

First, in the present work, we consider a non-dissipative system, such that $\gamma(t)$ and 
$g(t)$ are real. Next, the existence of dissipation due to three-body recombination processes
is also studied by changing the definition of $g(t)$ to a more general form, where the dissipation 
is parameterised by a constant $\kappa_3$, such that
\begin{equation}
g_c(t) = g(t) +{\rm i} \kappa_3 .
\label{eq04}
\end{equation} 
The different scenarios of the time modulations for the two- and three-body interactions can be 
exemplified by the following models:\\

\begin{enumerate}
\item[1.] {\bf Three-body interaction proportional to $[a_s(t)]^2$ (quadratic case).}

This case can occur in a model for a BEC with 1D non-polynomial GP equation,
confined in a cigar type trap~\cite{Salasnich}.
By a series expansion, valid for small $a_s|\psi|^2$, an effective quintic parameter can be
derived in Eq.~(\ref{eq01}), which is given by $G(t)\equiv 2\hbar\omega_{\perp}a_s^2(t)$.
A similar form of the corresponding equation, for a cigar-type trap, was also derived in \cite{Khaykovich}.
A quadratic dependence of $G(t)$ on $a_s(t)$ can also occur in the case when 
$\Gamma\equiv\Gamma(x,t) \approx \cos(\omega t)\cos(kx)$, corresponding to 
a time dependent short-scale nonlinear optical lattice. In this case,  averaged over short scale  
modulations in space, the dynamics is described by a GP equation with effective time dependent 
three body interactions\cite{SM,AAG0}. Another model with quadratic dependence on $a_s$ 
was also suggested in \cite{Tiesinga},  considering effective 3-body interactions for atoms 
loaded in a deep optical lattice.  \\

\item[2.] {\bf Three-body interaction proportional to $[a_s(t)]^4$ (quartic case).}

By varying $a_s(t)$ through Feshbach resonances techniques, as the absolute value of this 
two-body observable becomes very large, one approaches the unitary limit ($|a_s|\to\infty$) where 
many three-body bound-states and resonances can be found. This behaviour will induce changes in 
the corresponding quintic parameter of the GP equation, such that in Eq.~(\ref{eq01}) we have 
$G(t) \sim a_s^4(t)$\cite{Bulgac}.
\end{enumerate}

\section{Modulational instability }
\label{sec:cn}
In this section we consider a modulational instability (MI) of the nonlinear plane-wave solution 
for the Eq.(\ref{eq02}), such that
\begin{eqnarray}\label{eq05}
u_0\equiv u(0,t) &=& Ae^{{\rm i}\theta(t)}, \;\;\;{\rm where}\;\;
\theta(t) = A^2\int_0^t \left[\gamma(t') + A^2 g(t')\right] dt'.
\end{eqnarray}
To analyze MI we will look for a solution of the form
\begin{equation}
u(x,t) = [A + \delta u(x,t)]e^{{\rm i}\theta(t)},\;\;{\rm with}\;\;  \delta u \ll A.
\end{equation}\label{eq06}
By substituting the above expressions in Eq.(\ref{eq02}) and keeping only 
linear terms $\delta u\equiv \delta u(x,t) $,
 we have 
\begin{equation}\label{eq07}
{\rm i}\frac{\partial \delta u}{\partial t} + 
\frac{\partial^2 \delta u}{\partial x^2} + 
A^2 \left[\gamma(t) + 2A^2 g(t)\right](\delta u + \delta u^{\ast})=0.
\end{equation}
Now, by introducing $\delta u = v + {\rm i}w$, where $v\equiv v(x,t)$ and $w\equiv w(x,t)$,  and 
going to the corresponding Fourier components, $V\equiv V(k,t)$ and $W\equiv W(k,t)$,
 according to
\begin{equation}
\left(v,w\right) = \int e^{{\rm i}kx} \left(V,W\right) dk,
\label{eq08}\end{equation}
we obtain the system of equations:
\begin{eqnarray}
\frac{dV}{dt}  &=& k^2 W,\nonumber \\
\frac{dW}{dt} &=& - k^2 V + 2A^2 \left[\gamma(t) + 2A^2 g(t)\right]V.
\label{eq09}\end{eqnarray}
Finally we have
\begin{equation}
\frac{d^2V}{d t^2} + k^2\left[k^2  - 2A^2\left( \gamma(t) + 2A^2 g(t)\right)\right]V=0.
\label{eq10}\end{equation}

\subsection{Influence of the inelastic three-body collisions}

By taking into account inelastic three-body collisions, defined by a dimensionless parameter
$\kappa_3$, one should add the term ${\rm i} \kappa_3 |u|^4u$ in the Eq.(\ref{eq02}). 
In this case, by replacing $g(t)$ to $g_c(t)=g(t)+{\rm i}\kappa_3$, the Eqs.~(\ref{eq05}) and (\ref{eq06}) 
have to be replaced by
\begin{eqnarray}
u(x,t)&=&\left[A(t) + \delta u(x,t)\right] e^{{\rm i}\theta(t)},\;\;\;
A(t) = A_0 (1+ 4\kappa_3A_0^4 t)^{-1/4},\label{eq11}\\
\theta(t)&=&\int_0^t ds  \left[\gamma(s) A^2(s) + g(s)A^4(s)\right]\label{eq12}.
\end{eqnarray}
 In the above expression for $\theta(t)$, we neglect $\delta u(x,t)$ with the assumption 
that $A(t)\gg \delta u(x,t)$.
Next, by following the procedure of the previous subsection, with $\delta u = v + {\rm i} w$, 
for the Fourier component $V$ we obtain
\begin{eqnarray}
&&\frac{d^2V}{d t^2} + k^2\left[k^2  - 2A(t)^2( \gamma(t) + 2A(t)^2 g(t))\right] V = \nonumber\\
&& - 6\kappa_3 A(t)^4 \frac{dV}{dt} + 15\left[\kappa_3 A(t)^4\right]^2 V .
\label{eq13}\end{eqnarray}
 Therefore, due to inelastic three-body collisions, in Eq.~(\ref{eq10}) we have  the additional dissipative  
 term $6\kappa_3 A^4 dV/dt$, together with a term $\sim \kappa_3^2$, which can be neglected for 
  small $\kappa_3 $.
This will lead to the appearance of the threshold in the amplitude of modulations of the scattering 
length for the existence of the parametric resonances.

\subsection{Model of three-body interactions with quadratic dependence  on the scattering length}

\subsubsection{Modulational instability for periodic variations of the scattering length}

Next, in this subsection, we consider the MI for the case of  periodic modulations of the scattering length 
in time, given by $\gamma(t)$, with the three-body interaction term, $g(t)$, having a quadratic dependence 
on  $\gamma(t)$: 
\begin{equation}
\gamma(t) = \gamma_0 +\gamma_1\cos(\omega t), \ g(t) =  c\left[\gamma_0 + \gamma_1\cos(\omega t)\right]^2,
\label{eq14}\end{equation}
where $\gamma_0$ refers to the natural two-body scattering length, which can be attractive ($\gamma_0>0$)
or repulsive ($\gamma_0<0$); and $\gamma_1$ is the amplitude of the periodic modulation, such 
that we can take it as a positive quantity. 
We should note that Refs.~\cite{Gammal2000,Petrov} are mainly concerned with three-body repulsion 
as a way to stabilise a condensate with attractive two-body interaction. However, in the present case, as 
we are interested in examine the emergence of FW patterns, we consider to verify interesting conditions 
where the time-dependent parameter $g(t)$, given by Eq.~(\ref{eq14}), is positive ($c>0$), implying in 
attractive three-body interaction. 

By considering Eq.~(\ref{eq14}), without dissipation ($\kappa_3=0$),  from Eq.~(\ref{eq10}) we obtain
\begin{equation}
\frac{d^2V}{dt^2} + \Omega^2 \left[1 - f_1 \cos(\omega t) - f_2 \cos(2\omega t)\right]V =0,
\label{eq15}\end{equation}
where
\begin{eqnarray}
\Omega^2 &\equiv& k^2\Delta \equiv k^2 \left\{k^2 -2A^2\left[\gamma_0 +A^2c(2\gamma_0^2 + {\gamma_1^2})\right]\right\}, 
\label{eq16}\\
f_1 &\equiv& \frac{2\gamma_1 A^2(1+4cA^2\gamma_0)}{\Delta}, 
f_2 \equiv \frac{2c\gamma_1^2 A^4}{\Delta}.
\nonumber \end{eqnarray}
We have parametric resonances for two cases, at $\omega =2\Omega$ ($\eta\equiv 1$) and $\omega = \Omega$ 
($\eta\equiv 2$), such that the corresponding wavenumber $k_F^{(\eta)}$ is given by
\begin{eqnarray}
k_F^{(\eta)} &=& \pm \sqrt{ \frac{M_\pm}{2} + \frac{1}{2}\sqrt{M_\pm^2 + (\eta\omega)^2} } \equiv L_\eta ,\;\;\;
{\rm with}\nonumber\\ 
M_\pm&\equiv& 2A^2\left[\pm|\gamma_0| + A^2c\left(2\gamma_0^2 + {\gamma_1^2}\right)\right]
,\label{eq17}
\end{eqnarray}
where $M_+$ is for attractive or zero two-body interactions, $\gamma_0 \ge 0$, and $M_-$ for the repulsive case, 
$\gamma_0 <0$. In the present case, as we are analysing the case with $c>0$, $M_-$ can be set to zero or 
negative only for repulsive interactions. In the following, we consider only the relevant positive sign
for the resonance wavenumber $k_F$.

\begin{enumerate}
\item[1.] Let us consider more explicitly the  first resonance, $\omega = 2\Omega$ ($\eta=1$). \\

In the attractive or zero two-body interactions, $\gamma_0 \ge 0$, $M_{+}>0$, the wavenumber $k_F^{(1)}$
(corresponding to a length $L_1$) of the Faraday pattern, is such that
\begin{equation}
k_F^{(1)}\equiv\frac{2\pi}{L_1}=\sqrt{\frac{M_{+}}{2}\left(\sqrt{1+\frac{\omega^2}{M_{+}^2}} +1\right)},
\label{eq18}\end{equation}
where $L_1$ gives the period of the generated Faraday pattern in {\it space}.
For large frequencies of modulations, with $\omega\gg M_+$, the period behaves like
$ L_1 \sim {1}/{\sqrt{\omega}}.$

In the  case of repulsive two-body interactions, $\gamma_0 <0$,  which can be more easily explored in
experiments, we have two possibilities, with $M_-$ positive or negative. When $M_-$ is positive, we can 
use the same expression for $k_F$ as Eq.~(\ref{eq18}) with $M_{+}$ 
replaced by $M_{-}$. However, for such repulsive interaction, $M_-<0$ can only be satisfied
if  $\gamma_1 < {1}/{(2\sqrt{2}cA^2)}$ and $|\gamma_0|$ is within the interval
\begin{eqnarray}
1+\sqrt{1-2(2cA^2\gamma_1)^2}>(4cA^2|\gamma_0|)> 1 -\sqrt{1-2(2cA^2\gamma_1)^2}.
\label{eq19}\end{eqnarray}
If these conditions are satisfied, the Faraday patterns are given by 
\begin{equation}
k_{F_-}^{(1)} \equiv\frac{2\pi}{L_1}= \sqrt{\frac{|M_{-}|}{2}\left(\sqrt{1+ \frac{\omega^2}{M_{-}^2}}-1\right)}.
\label{eq20}\end{equation}
 However, we should noticed that the above conditions with (\ref{eq19}) are too much restrictive
for experimental observation. Instead, in the repulsive case ($\gamma_0<0$), one should search 
FW patterns outside this limit, when Eq.(\ref{eq18}) can be applied. In Fig.~\ref{f01}, we show the 
behaviour of the period of the oscillations as a function of the amplitude $\gamma_1$, for two cases
of repulsive interactions, with fixed $\gamma_0 =$ -0.5 and -1.
\begin{figure}[tbph]
\centerline{
\includegraphics[width=6.cm,clip]{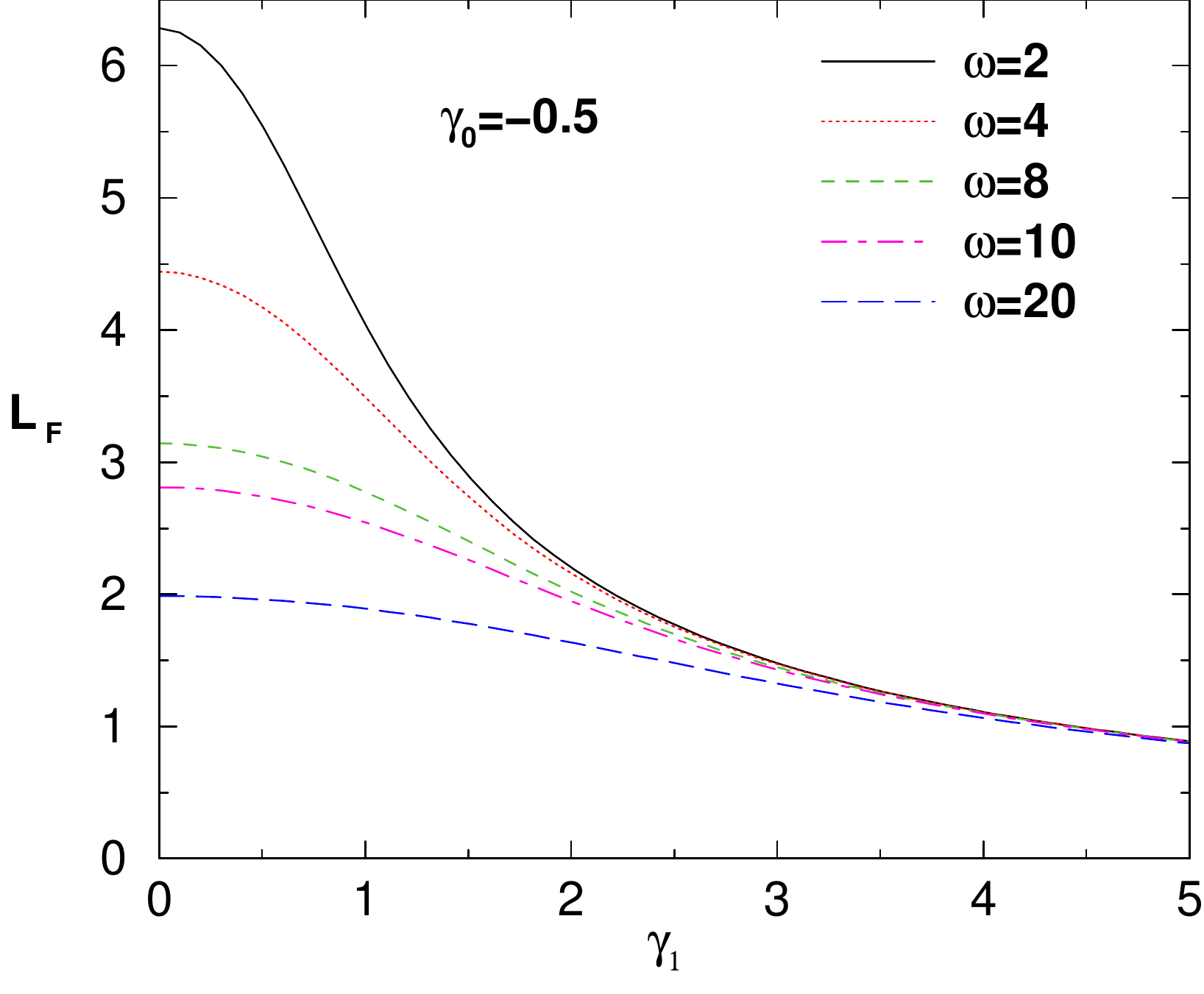}
\hspace{0.1cm}
\includegraphics[width=6.cm,clip]{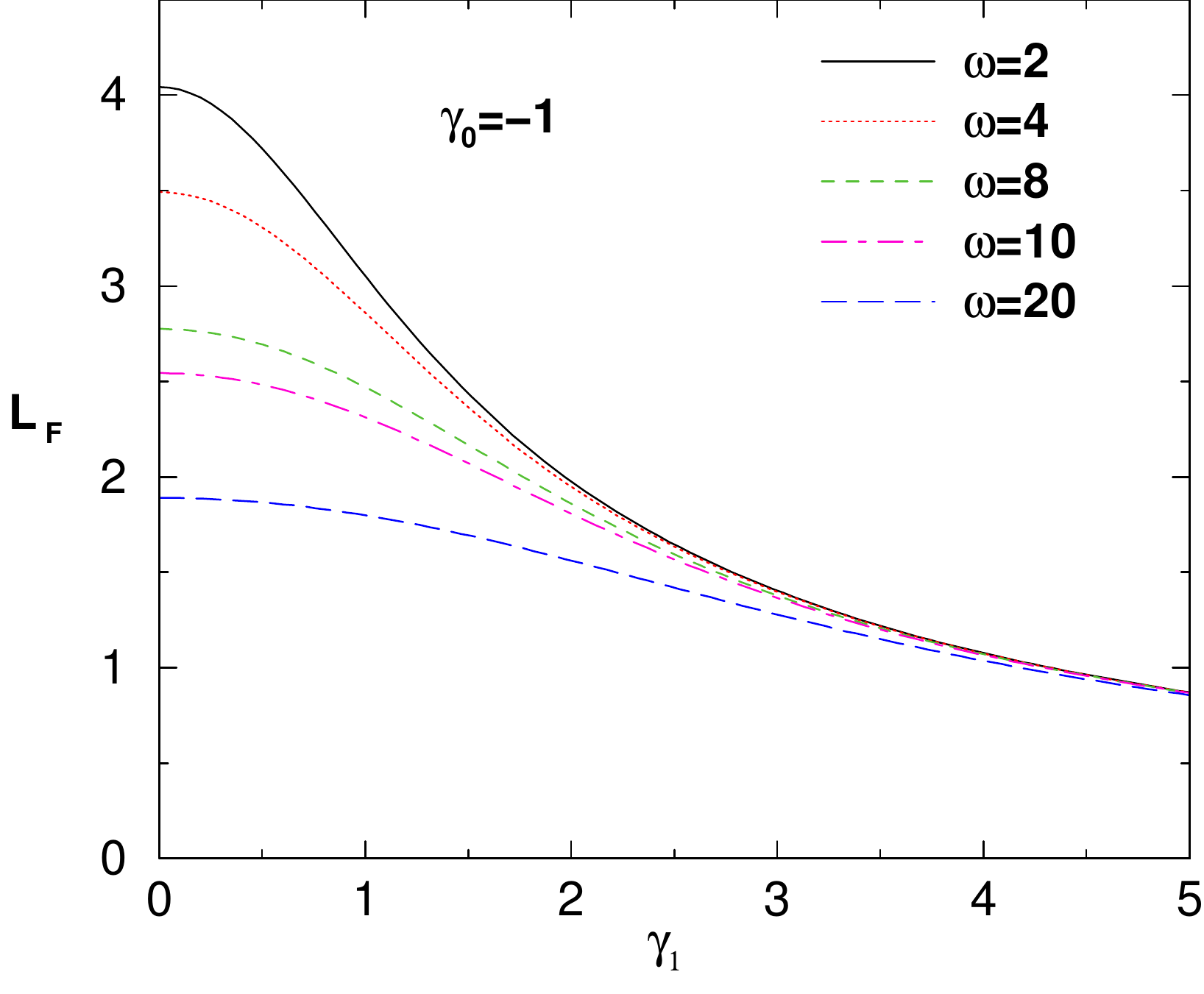}}
\caption{(color on-line) 
When the three-body interaction is proportional to $[a_s(t)]^2$, given by Eq.~(\ref{eq14}), we 
show the behaviour of the period of FW oscillations, $L_F$ (= $L_1$ 
in case $2\Omega=\omega$, and = $L_2$, when $\Omega=\omega$), given as functions of 
$\gamma_1$, for a few set of frequencies $\omega$ and for two cases of two-body
repulsive interactions. All quantities are dimensionless and we fix the other parameters such 
that $A=c=1$.
}\label{f01}
\end{figure}
\item[2.] For the second parametric resonance, $\omega =\Omega$ ($\eta=2$), the corresponding pattern is only due to the three-body effects.
In this case, by following  Eq.~(\ref{eq17}), we obtain the same expressions (\ref{eq18}) and (\ref{eq20}) for $\eta=2$, 
as in the first case shown above, with replacement of $\omega$ by $2\omega$.
For the period, we have  $ L _1 = \sqrt{2}L_2$.
\end{enumerate}

From Eqs. (\ref{eq18}) and (\ref{eq20}), we observe the existence of an additional dependence 
on the wavenumber of Faraday pattern from the amplitude of modulations $\gamma_1$. 
This result is new, as far as we know, since in previous 
investigations~\cite{LVS,Engels,Nicolin2} $k_F$ is independent on $\gamma_1$. For large 
$\gamma_1 \gg 1$ we have estimated that  $L_F \sim 1/\gamma_1 $. Thus, by varying $\gamma_1$  
and with the knowledge of the effective parameters for the two- and  three-body interactions, one can
 tune the corresponding period of the Faraday pattern.

\subsubsection{Modulational instability for fast periodic variations of the scattering length.}

Let us consider the case of strong fast modulations, when $\omega, \gamma_1 \gg 1$ and 
$\gamma_1/\omega \sim O(1)$.
Following Refs.~\cite{AG,Jar},  it is useful to perform the following change of variables:
\begin{equation}\label{eq21}
u(x,t)= U(x,t)\exp\left({{\rm i}\Gamma_1(t) |U(x,t)|^2 + {\rm i} \Gamma_2(t) |U(x,t)|^4}\right),
\end{equation}
where $\Gamma_1(t)$ and $\Gamma_2(t)$ are antiderivatives of $\gamma_1(t)$ and $g(t)$, respectively:
\begin{equation}
\Gamma_1(t) = \int_0^t ds\; \gamma_1(s)\;\;{\rm and}\;\; \Gamma_2(t) = \int_0^t ds\; g(s).
\label{eq22}\end{equation}
$U\equiv U(x,t)$ is a slowly varying function of $x$ and $t$. To find out the GP equation, averaged 
over period of fast oscillations, we first obtain the averaged Hamiltonian. By substituting 
the Eq.(\ref{eq21}) into the expression for the Hamiltonian and by averaging over
the period of rapid oscillations, we obtain
{\small
\begin{eqnarray}\label{eq23}
&&H_{av}=\int_{-\infty}^{\infty}dx{\cal H}_{av}\nonumber\\
&&{\cal H}_{av}=\left\{\left|\frac{dU}{dx}\right|^2 +\left[
\sigma_1^2\left(\frac{d|U|^2}{dx}\right)^2 + \sigma_2^2
\left(\frac{d|U|^4}{dx}\right)^2\right]|U|^2 - \frac{g_{eff}}{3}|U|^6\right\},
\\
&&{\rm where}\;\;
\sigma_{1}^2 \equiv \frac{c\gamma_1^2}{2\ \omega^2}, \;\; 
\sigma_2^2 \equiv \frac{c\gamma_1^4}{32\ \omega^2}, 
\ g_{eff} \equiv c\frac{\gamma_1^2}{2}. 
\nonumber \end{eqnarray}}
The averaged GP equation is given by
{\small 
\begin{eqnarray}
&&{\rm i}\frac{\partial U}{\partial t}=\frac{\delta {\cal H}_{av}}{\delta U^{\ast}}\nonumber\\
\label{eq24}
&&=-\frac{d^2U}{dx^2}-(\sigma_1^2 + 16\ \sigma_2^2 |U|^4)
\left[
\left(\frac{d|U|^2}{dx}\right)^2
+ 2|U|^2 \frac{d^2 |U|^2 }{dx^2}\right]U  
- g_{eff}|U|^4 U.
\end{eqnarray}
}

Performing standard MI analysis with the averaged equation (\ref{eq24}), i.e. considering the evolution of 
the perturbed nonlinear plane wave solution in the form:
\begin{equation}
U = (A + \delta U(x,t))e^{{\rm i}g_{eff}A^4 t}.
\label{eq25}\end{equation}
{Next, with $\delta U\equiv P + {\rm i}Q$, by  performing the corresponding Fourier transforms
[$P=\int dk p(k) \exp({\rm i}kx)$ and $Q=\int dk q(k) \exp({\rm i}kx)$],
we obtain the dispersion relation
\begin{equation}
\Omega_{av}^2 = k^2[(1+r)k^2 - 4g_{eff}A^4], \;\;{\rm where}\;\; r\equiv 2A^4 (\sigma_1^2 + 4\sigma_2^2 A^4).
\label{eq26}\end{equation}
From this expression, the maximum gain occurs at the wavenumber 
\begin{equation}
k_c = \frac{\sqrt{2g_{eff}}A^2}{\sqrt{1+r}}, 
\label{eq27}\end{equation}
with the corresponding gain rate given by 
\begin{equation}
p_c = \frac{2\sqrt{g_{eff}}A^4}{\sqrt{1+r}}.
\label{eq28}\end{equation}
Note that management of the scattering length can suppress the MI and will correspond to make weaker the 
effective three-body interactions. This effect leads to arrest of collapse  and make possible the existence of 
stable bright matter wave solitons in BEC with effective attractive quintic nonlinearity (see also Ref.~\cite{Mur}).

The procedure by averaging out two body processes, with enhancement of the three-body (attractive) 
interactions, in the quasi one dimensional geometry, can lead to the collapse, which will happen when 
the number of atoms $N$ exceed a critical value $N_c$. However, for the strong and  rapid modulations case, 
as we verify from the averaged equation (\ref{eq24}), a nonlinear dispersion term appears.
This term showed  that for small widths the effective repulsion can arrest the collapse, such that stable bright 
solitons can exist for $N>N_c$. This problem requires a separate investigation.
 
\subsection{The model of three-body interactions with quartic dependence on the scattering length}
Let us now consider the case  when the strength of  three-body interactions is proportional to quartic power of the scattering length; i.e., when $g(t) \sim a_s^4$~\cite{Bulgac}, such that
\begin{equation}
\gamma(t) = \gamma_0 + \gamma_1\cos(\omega t), \;\;\;  
g(t)=c_E \left[\gamma_0 + \gamma_1\cos(\omega t)\right]^4.
\label{eq29}\end{equation}
The possibility of attractive or repulsive three-body interaction will correspond, respectively, to $c_E$ being 
positive or negative, which can happen for attractive or repulsive two-body interactions. In the next, in our 
numerical search for the FW patterns near the Efimov limit, both the cases are verified.

From Eq.~(\ref{eq13}) without the term $\sim\kappa_3^2$,  the expression for 
$V\equiv V(k,t)$ 
is given by
\begin{equation}
\frac{d^2V}{d t^2} + \Omega_1^2\left[1-\sum_{j=1}^4 h_j\cos(j\omega t)\right]V + 6\kappa_3 A^4 \frac{dV}{dt}=0,
\label{eq30}\end{equation}
where
\begin{eqnarray}
\left.
\begin{array}{ll}
\Omega_1^2&\equiv k^2(k^2 - 2A^2 a_0),\\ 
h_j&\equiv \displaystyle  \frac{a_j}{k^2-2A^2 a_0}\;\; (j=1, 2, 3, 4),\\
a_0 &\equiv \gamma_0 + 2c_E A^2\left[\gamma_0^4 + 3\gamma_0^2\gamma_1^2 + 
({3}/{8})\gamma_1^4\right],\\
a_1&\equiv 2A^2\gamma_1\left[
1 + 2c_E A^2\left(4\gamma_0^3 + 3\gamma_0\gamma_1^2\right)\right],\\ 
a_2&\equiv 2c_E A^4 \gamma_1^2\left(6\gamma_0^2 + \gamma_1^2\right),\\
a_3&\equiv 4c_E A^4 \gamma_0\gamma_1^3,\\
a_4&\equiv \displaystyle  (c_E/2) A^4 \gamma_1^4.
\end{array}
\right\}
\label{eq31}\end{eqnarray}
We should also noticed that in the above equations, for $\kappa_3\ne 0$, we have $A\equiv A(t)$, 
given by
Eq.~(\ref{eq11}), such that 
$h_j\equiv h_j(t)$ and $a_j\equiv a_j(t)$.

The parametric resonances occur at
\begin{equation}
\frac{\eta\omega}{2}=\Omega_1\;\;\; (\eta=1, 2, 3, 4).
\label{eq32}\end{equation}
The first parametric resonance, given by $\eta=1$, occurs for 
$
{\omega}=2(\Omega_1 + \Delta),
$
where $\Delta$ is detuning.
Note that, as expected, all the resonance are absent when $\gamma_1 =0$, as one can verify in the
above expression, where  $h_j=0$. 

When $a_0 >0$, which can happen for attractive as well as for repulsive two body interactions, 
we obtain
\begin{equation}
k_F = \sqrt{a_0}A\sqrt{1 + \sqrt{1 + \frac{\omega^2}{4A^4a_0^2}}}.
\label{eq33}\end{equation}
And, in the case that $a_0 < 0$,  which can occur for repulsive two-body interactions ($\gamma_0<0$)
when $c_E>0$, as well as for attractive two-body interactions ($\gamma_0>0$) if $c_E<0$, the
wavenumber is given by
\begin{equation}
k_F =  \sqrt{|a_0|}A\sqrt{ \sqrt{1 + \frac{\omega^2}{4A^4a_0^2}}-1}.
\label{eq34}\end{equation}
Again we observe that the FW pattern, given by $k_F$, will depend on the modulation amplitude of 
the scattering length, $\gamma_1$, in view of the expression for $a_0$ given in (\ref{eq31}). In the 
limit of large values for this amplitude and negative $c_E<0$ corresponding to the repulsive three-body interactions, 
we have $k_F \sim \gamma_1^2$, such that the FW period 
will be given by  $L_F \sim 1/\gamma_1^2$. 

For small modulation amplitudes, $h_1$, and for  three-body losses $\kappa_3$,  we can perform 
the analysis based on the perturbation theory.  The boundary value for instability of detuning,
$\Delta_c$, and the corresponding parametric gain $p_{max}$, are given by
\begin{equation}
\Delta_c = \frac{\sqrt{h_1}}{4\Omega_1}\;\;\;{\rm and}\;\;\;
p_{max}=\frac{\sqrt{h_1}\omega}{2}.
\label{eq35}\end{equation}
The threshold value of the amplitude of modulations  when the resonance occurs can be found from the 
condition  $h_1 (k=k_{max}) = 6\kappa_3 A^4 $. By taking into account that $k_{max}^2 = a_0A^2$,  
considering $\gamma_1 \ll 1$, and neglecting the terms $\sim  \kappa_3^2, \gamma_1^2$  we obtain
\begin{equation}
\gamma_{1,th} =\frac{3\kappa_3 A^4 a_0}{1+2c_E A^2\gamma_0^3}.
\label{eq36}\end{equation}
In numerical simulations, with $A=\gamma_0=c_E =1$, leads to $\gamma_{1,th} =1.8 \kappa_3$. 
We can look for  $\kappa_3=0.01,\;\; 0.03$.

The next resonance occurs at $\eta=2$, with $\omega =\Omega_1 + \Delta_1 $, with the wavenumber
\begin{equation}
k_F = \sqrt{a_0}A\sqrt{1+\sqrt{1+\frac{\omega^2}{A^4a_0}}}.
\label{eq37}\end{equation}
In the particular case of $\gamma_0=0$, we have $a_0=a_0^0=3 c_E\gamma_1^4/8$
and
\begin{equation}
k_F^{0} = k_F(a_0^0).
\label{eq38}\end{equation}
In this case, the boundary value of instability of detuning $\Delta_{1c}$ and the corresponding
parametric gain $p_{max}$ are given by
\begin{equation}
\Delta_{1c} = \frac{\sqrt{h_2}}{2\Omega_1}\;\;\;{\rm and}\;\;\;
p_{max}=\frac{\sqrt{h_2}\omega}{2}.
\label{eq39}\end{equation}
The perspective on studying FW resonances with three-body parameter having a quartic dependence 
on the $s-$wave two-body scattering length $a_s$, can happen when  $a_s$ is negative (unbound two-body states) 
and very large, near the Efimov limit (where the number of three-body bound states, as well as three-body 
resonances, are expected to increase as the system approaches the unitary limit  $|a_s|\to\infty$)~\cite{Braaten2003}. 
Near this limit, the three-body parameter goes with a fourth power of $a_s$ and can be positive or negative. 
The corresponding contribution of three-body collisions to the ground-state energy is proportional to the three-body 
parameter and density ($\sim g\, n^3$). In this case, for repulsive $g$, one can find a stable ground state~\cite{Bulgac}.
 In Fig.~\ref{f03}, it is shown the  evolution of the central densities for the second parametric resonance,
which occurs at the theoretical predicted value, $k=4.5$. 

\section{Numerical simulations}
\label{sec:stab}
 First, by considering the case when the strength of the three-body interaction is related
to the square of the two-body scattering length $a_s$, we present our results in three figures,
Figs.~\ref{f02}, \ref{f03} and \ref{f04}.
The emergence of a parametric resonance is displayed in Fig.~\ref{f02}, for some specific 
dimensionless parameters, with the modulated two-body parameter given by $\gamma_0=0$, 
$\gamma_1=0.5$ and $\omega=20$. The amplitude $A$ and three-body parameter $c$
are fixed to one, as indicated in the respective captions. The results were obtained for $(1-|u|^2)$ 
in the central position, as a function of time, by varying the wave number $k$.

\begin{figure}[tbph]
\centerline{
\includegraphics[width=7.cm,clip]{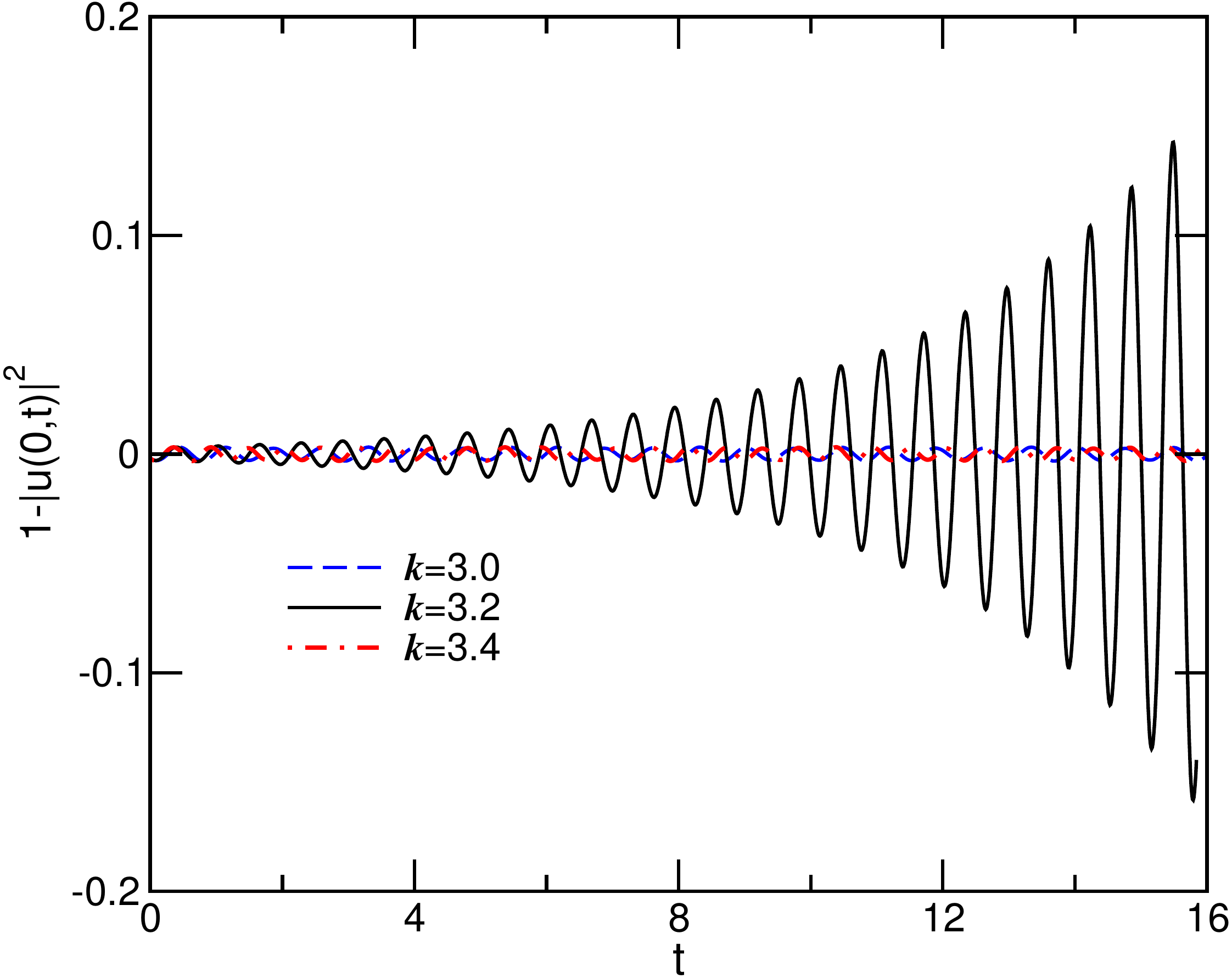}\hspace{0.1cm}
\includegraphics[width=7.cm,clip]{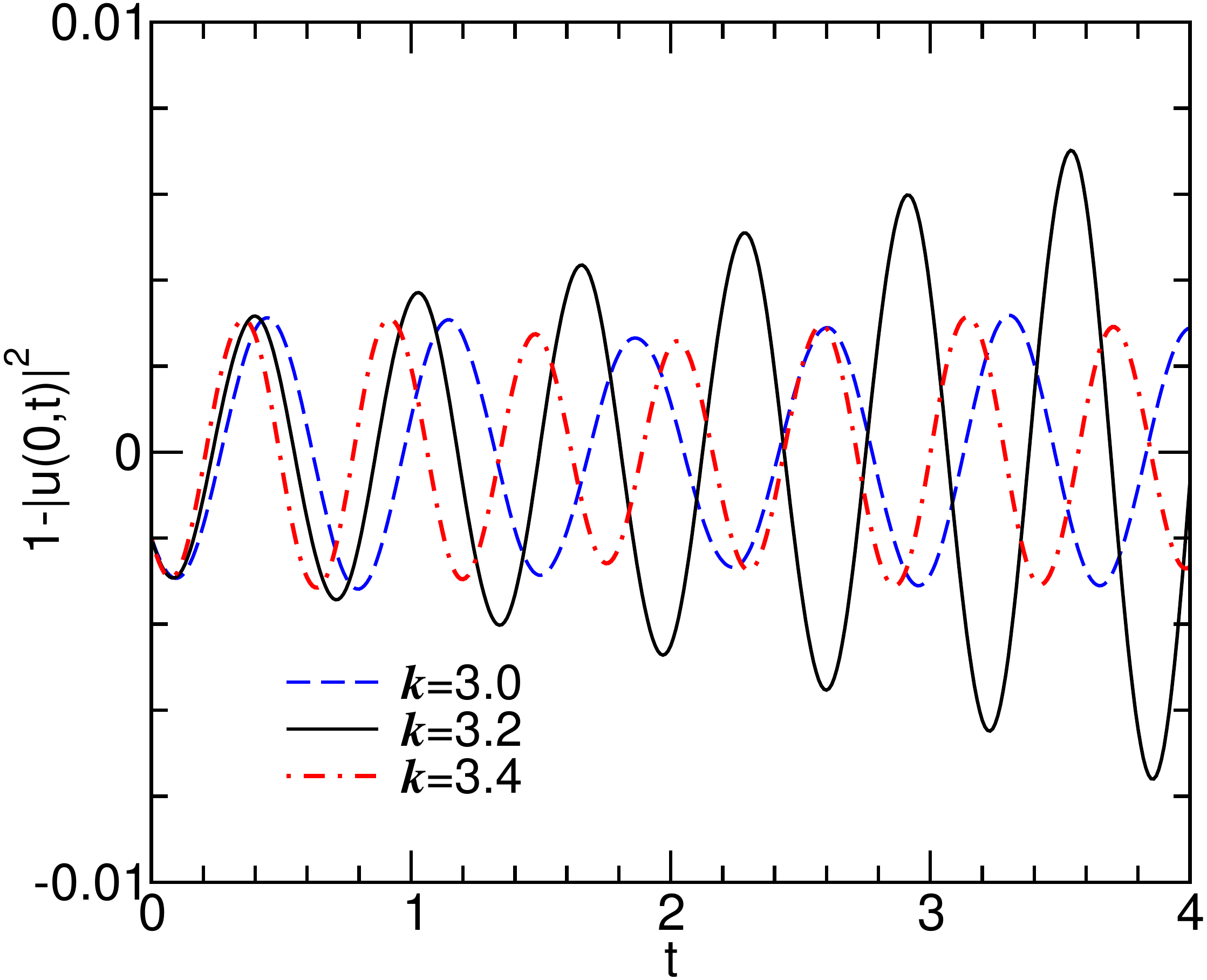}
}
\caption{(color on-line) Behavior of the central density $|u(0,t)|^2$, as function of time, showing the 
emergence of the first parametric resonance (for $\omega=20$), from full-numerical calculations.
In full agreement with analytical predictions for the values of $k$, it is
obtained the resonance for $k=k_F=3.2$. The other parameters, in this case, are such 
that $\gamma_0=0$, $\gamma_1=0.5$, $\eps_0=0.001$, $A=1$, and $c=1$, with all 
quantities in dimensionless units. In the right frame, we show a smaller time interval ($t<4$)
for a clear identification of the plots for the given values of $k$.}
\label{f02}
\end{figure}
\begin{figure}[tbph]
\centerline{
\includegraphics[width=7.cm,clip]{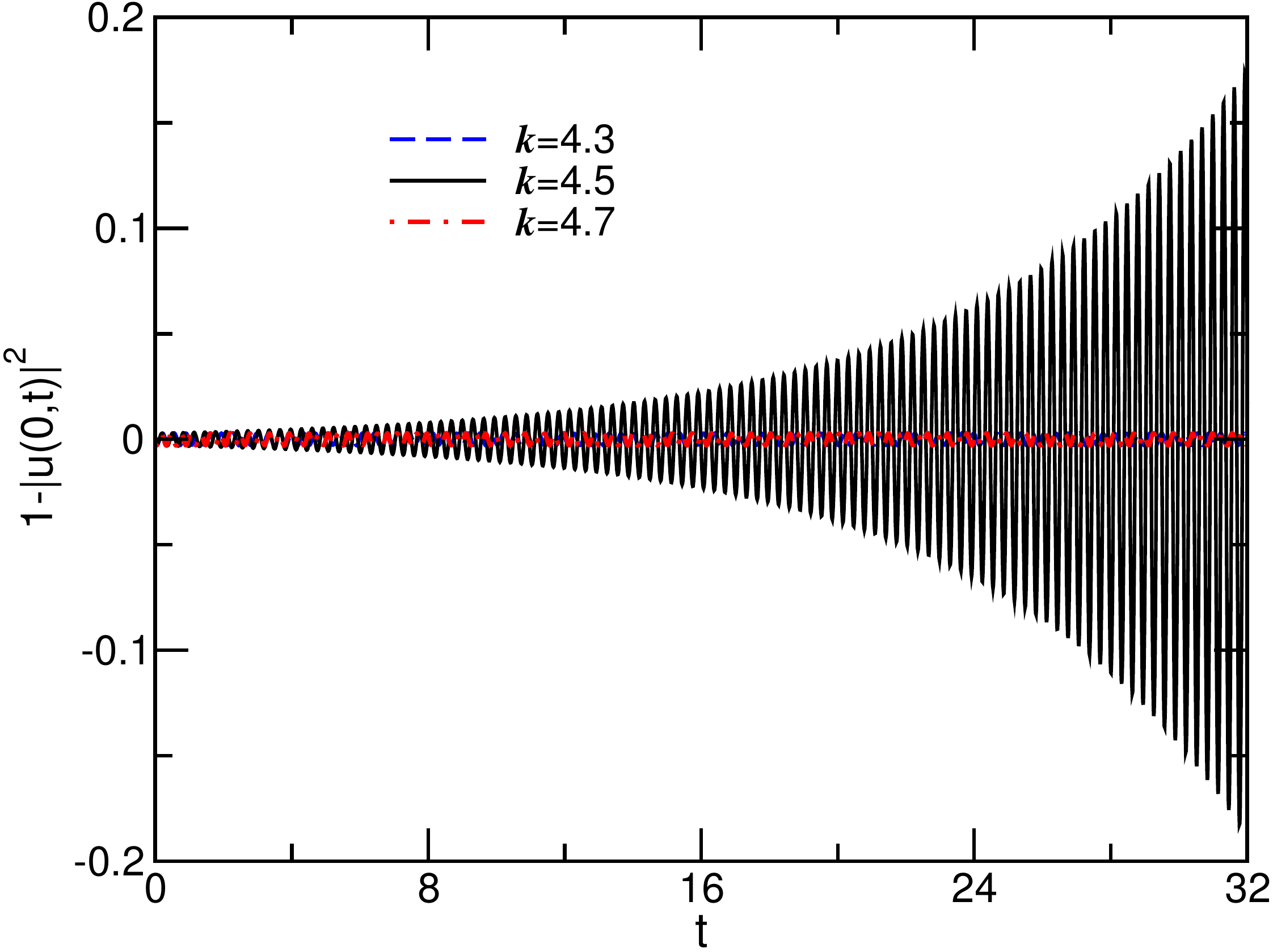}
\hspace{0.1cm}
\includegraphics[width=7.cm,clip]{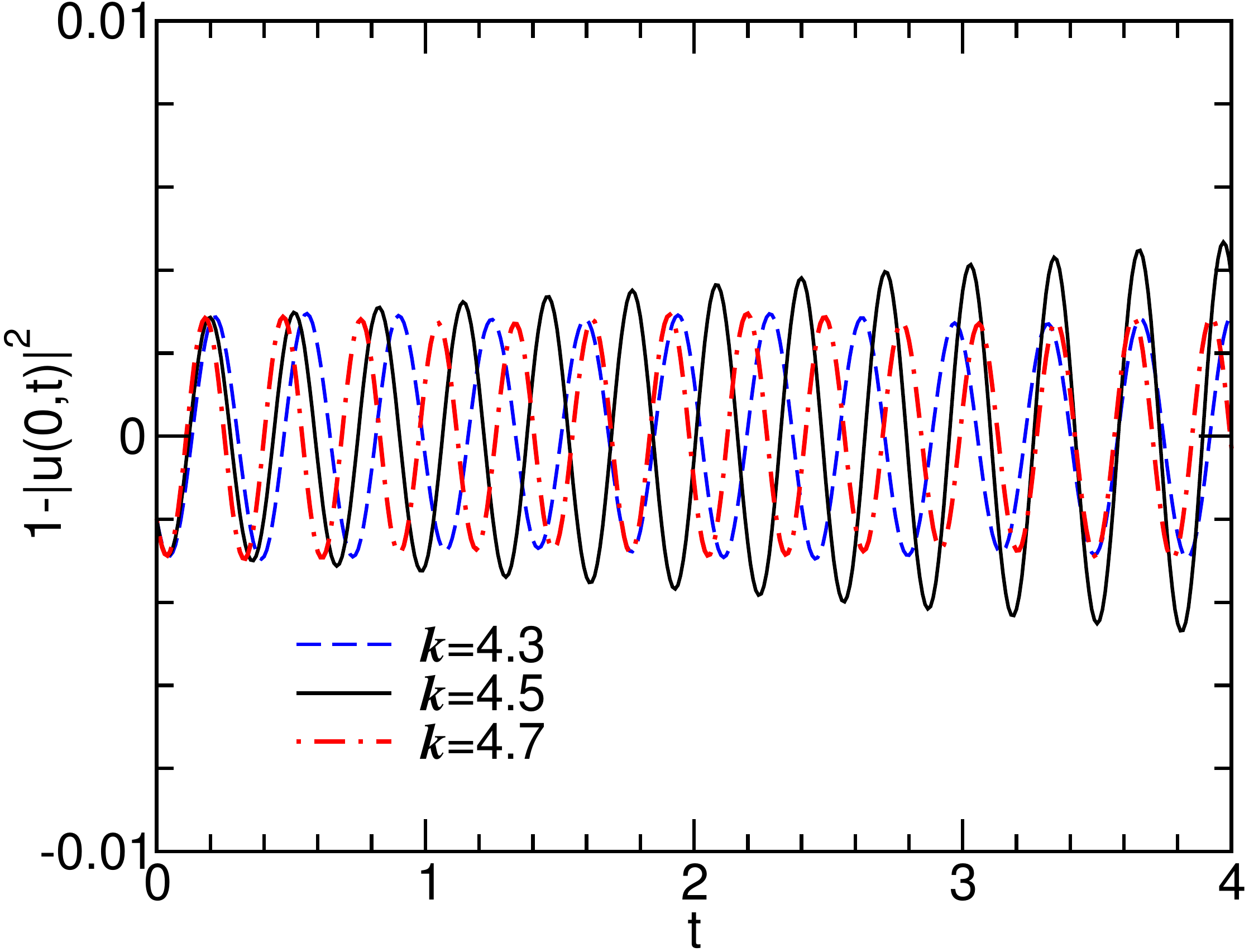}}
\caption{(color on-line) 
Following Fig.~\ref{f02}  the behavior of central density is displayed as a function of time, 
showing the emergence of the second parametric resonance ($\omega=40$), from 
full-numerical calculations. In this  case, again in agreement with analytical prediction, 
the resonance occurs at $k=4.5$. In the right frame, for $t<4$, we also show the plots for the 
given $k$, in order to appreciate how the resonance starts to appear.
The other parameters are the same as in  Fig.~\ref{f02}.
}
\label{f03}
\end{figure}
\begin{figure}[tbph]
\centerline{
\includegraphics[width=8cm,clip]{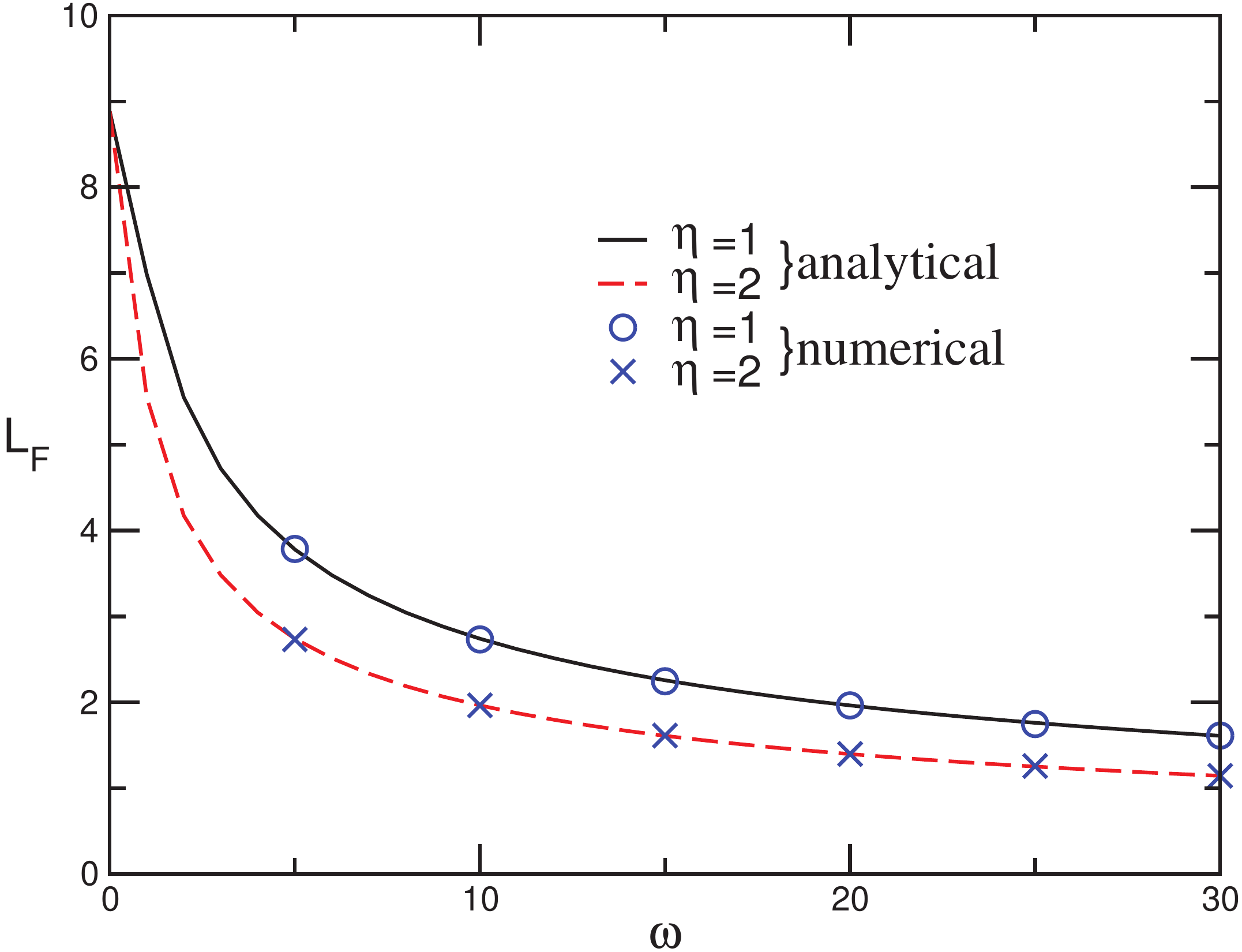}
}
\caption{(color on-line) The length of Faraday pattern, $L_F$, is presented as 
a function of the frequency, for the first ($\eta=1$) and second ($\eta=2$) resonances,
by considering analytical (solid and dashed lines) and numerical results (empty circle
and crosses). As shown, we have a perfect agreement between analytical and numerical 
results. $L_F$ and  $\omega$ are in dimensionless units.}
\label{f04}
\end{figure}

The dependence of the spatial period of the Faraday pattern, $L_F$, on the frequency of modulations is 
presented in Fig.~\ref{f04}, for the first and second resonance. As shown, the analytical predictions 
given by Eqs.~(\ref{eq18}) and (\ref{eq20}) are in good agreement with full numerical calculations.  

By considering the second model, when the strength of the three-body interactions is proportional to the 
fourth power of of the two-body scattering length, we present the results of numerical simulations  in 
Figs.\ref{f05}-\ref{f11}. As we have considered in the first case, given in Figs. \ref{f02} and 
\ref{f03},  for this second case the evolution of the central density with the time are plotted in 
Figs. \ref{f05} and \ref{f06}, considering the first and second parametric resonance, respectively.
The theoretical predictions for the positions of the resonances are quite well reproduced by the 
numerical results. The grow rate for the second resonance is shown to be much  slower than for 
the first one. 

The influence of  dissipation, due to inelastic three-body interactions on the process of the Faraday 
pattern generation, is presented in Fig.~\ref{f07}, considering the case for the resonance value
 $k=3.17$, which was shown (without dissipation) in Fig.~\ref{f05}. It is observed that the amplitude of the resonance 
 decreases gradually with increasing of the dissipation, as expected. In this figure, the value of the dissipation parameter,
 $0<\kappa_3<1$, is presented inside the frame close to the corresponding plot.  
 In Fig.~\ref{f08}, the observed results are demonstrating  the existence of a threshold in the amplitude of the
 modulations, given by $\gamma_1$. To verify that, we have selected from Fig.~\ref{f07} the case of $\kappa_3 =0.025$. 
 
In case of repulsive interactions ($\gamma_0<0$), we first present modelling results for the first and second 
parametric resonance in Figs.~\ref{f09} and \ref{f10}, by considering that the three-body parameter is attractive,
such that $c_E>0$ in Eq.~(\ref{eq29}). 
The corresponding predicted values, $k=3.14$ (first resonance) and $4.46$ (second resonance), are confirmed 
by the numerical simulations. The growth rate, for the second resonance, again goes slower than the case of the 
first parametric resonance. The results, in both the cases (first and second resonance) are compared with two 
values of $k$ outside of the position of the resonance. In second panels of these Figs.~\ref{f09} and \ref{f10}, as
in some of our previous results, we show small intervals in time of the respective results shown in the
first panels, for better identification of the plots.

As a final result, in Fig.~\ref{f11}, we found useful also to present one model result for the quartic case, with repulsive 
two-body interaction, when the three-body interaction is also repulsive, such that $c_E<0$ in Eq.~(\ref{eq29}). 
In this case, only the first parametric resonance is shown, with $\omega=20$ and $c_E=-1$, considering that 
the resonant position is very well defined by the analytical expression, $k_F=3.129$, with results similar to the 
ones presented in Fig.~\ref{f09},  such that the second resonance position can be easily predicted by using the 
corresponding analytical expression. As shown, the resonant pattern grows faster in case of repulsive three-body 
interaction.

In the above full-numerical results presented in this work, the simulations were performed by using split-step 
fast-Fourier-transform (FFT) algorithm, with boundary conditions enough extended to avoid reflection effects 
on the evolutions. In order to facilitate the emergence of Faraday patterns, we started from a uniform density 
profile of modulus one, adding a small perturbation with the form  $u_{initial}=1+\eps_0 (1+{\rm i})\cos(kx)$. 
The results obtained for the evolutions of the density profiles are quite stable numerically, such that one can
easily verify the resonance positions.  As a final remark on the present numerical approach, supported by our 
comparison of results obtained for Figs.~\ref{f09} and \ref{f11}, in the quartic case with repulsive two-body 
interaction, it is worthwhile to point out that we found the resonant behaviour more 
stable for longer evolution times when considering repulsive three-body interactions than in the case with 
attractive three-body interactions.

\begin{figure}[tbph]
\centerline{
\includegraphics[width=7.cm,clip]{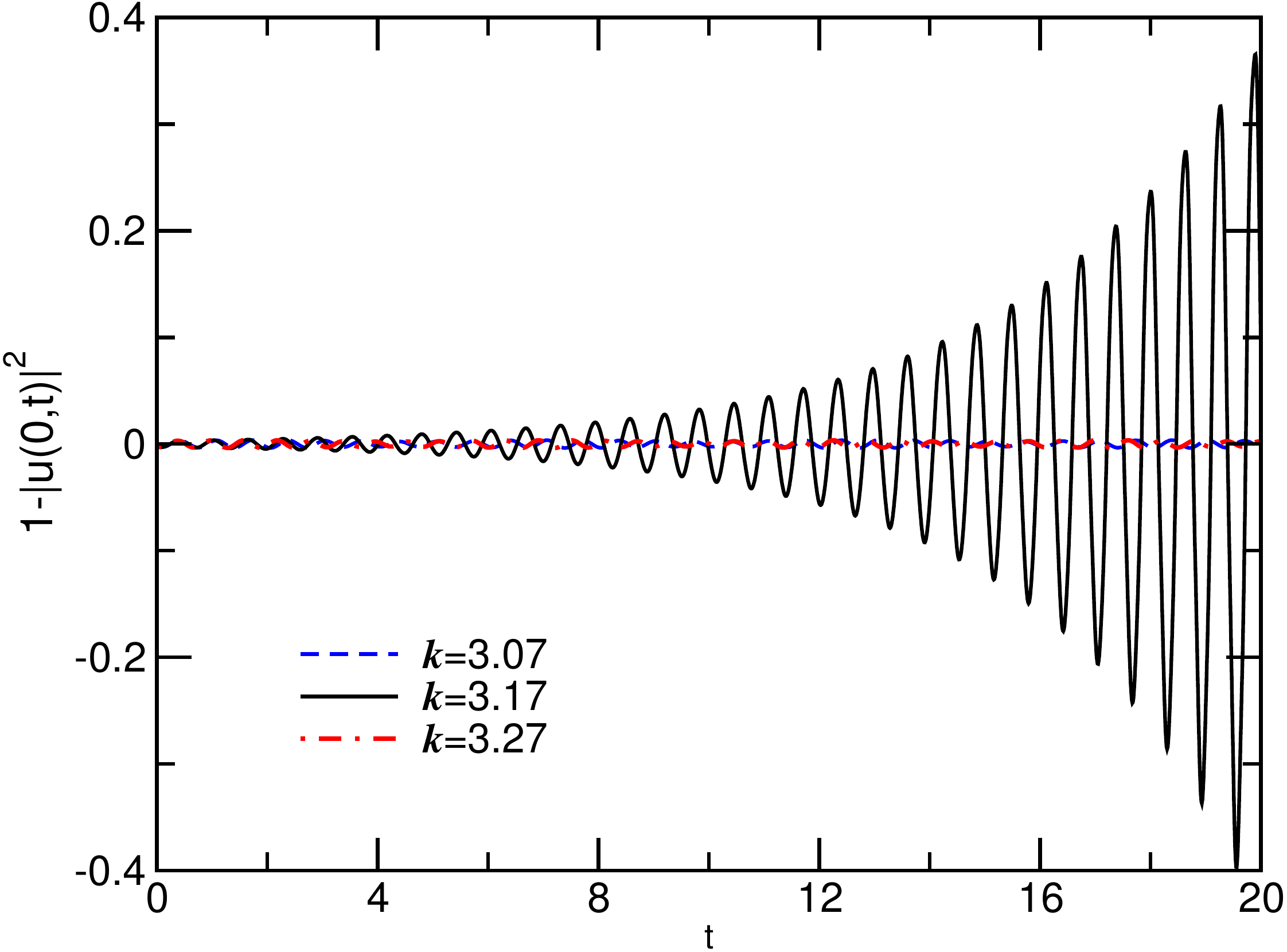}
\hspace{0.1cm}
\includegraphics[width=7.cm,clip]{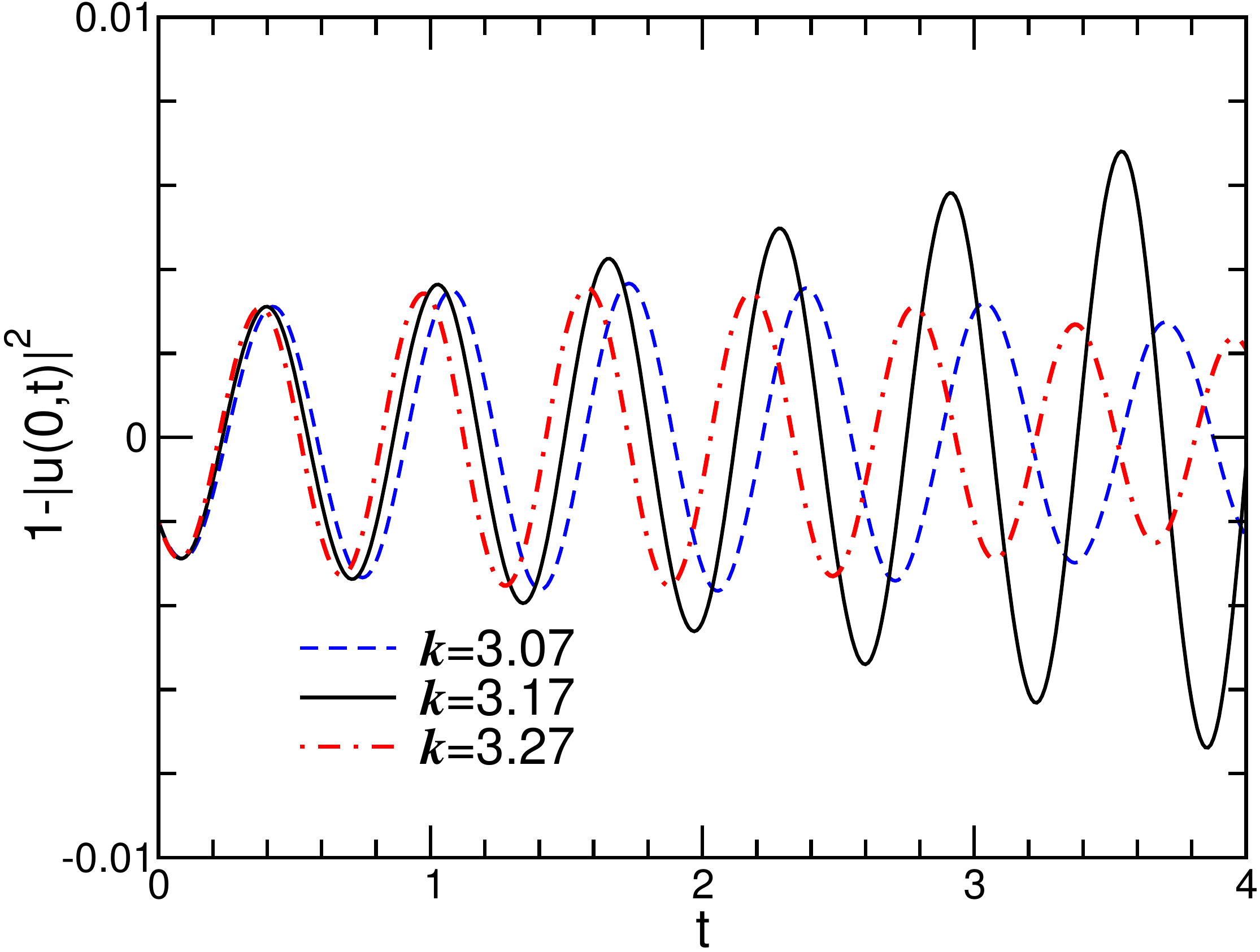}}
\caption{(color on-line) For the case when the behavior of three-body interactions goes with the
power four of the two-body interactions, in analogy with Figs.\ref{f02} and 
\ref{f03}, we present the behavior of central density, as function of time, 
showing the emergence of the first parametric resonance (for $\omega=20$), 
from full-numerical calculations. As before, we follow the analytical prediction, obtaining the 
resonance for $k=3.17$. As in previous figures, the other parameters, in 
dimensionless units, are such that  $\gamma_0=0$, $\gamma_1=0.5$, $\eps_0=0.001$, $A=1$, and $c=1$.  
We also present a plot, in the right frame, considering a smaller interval $t<4$.}
\label{f05}
\end{figure}
\begin{figure}[tbph]
\centerline{\hspace{0.5cm}
\includegraphics[width=7.cm,clip]{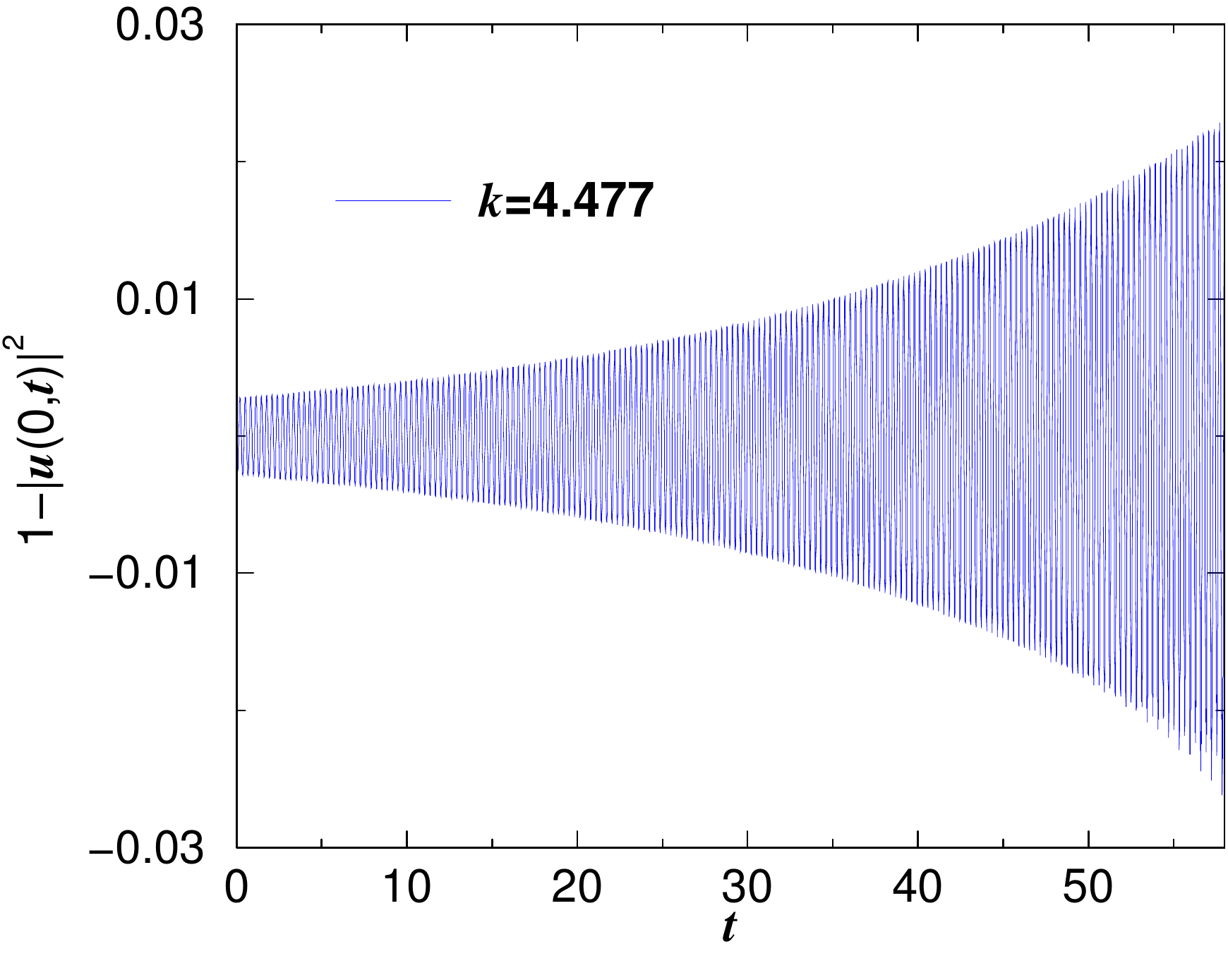}
\includegraphics[width=7.cm,clip]{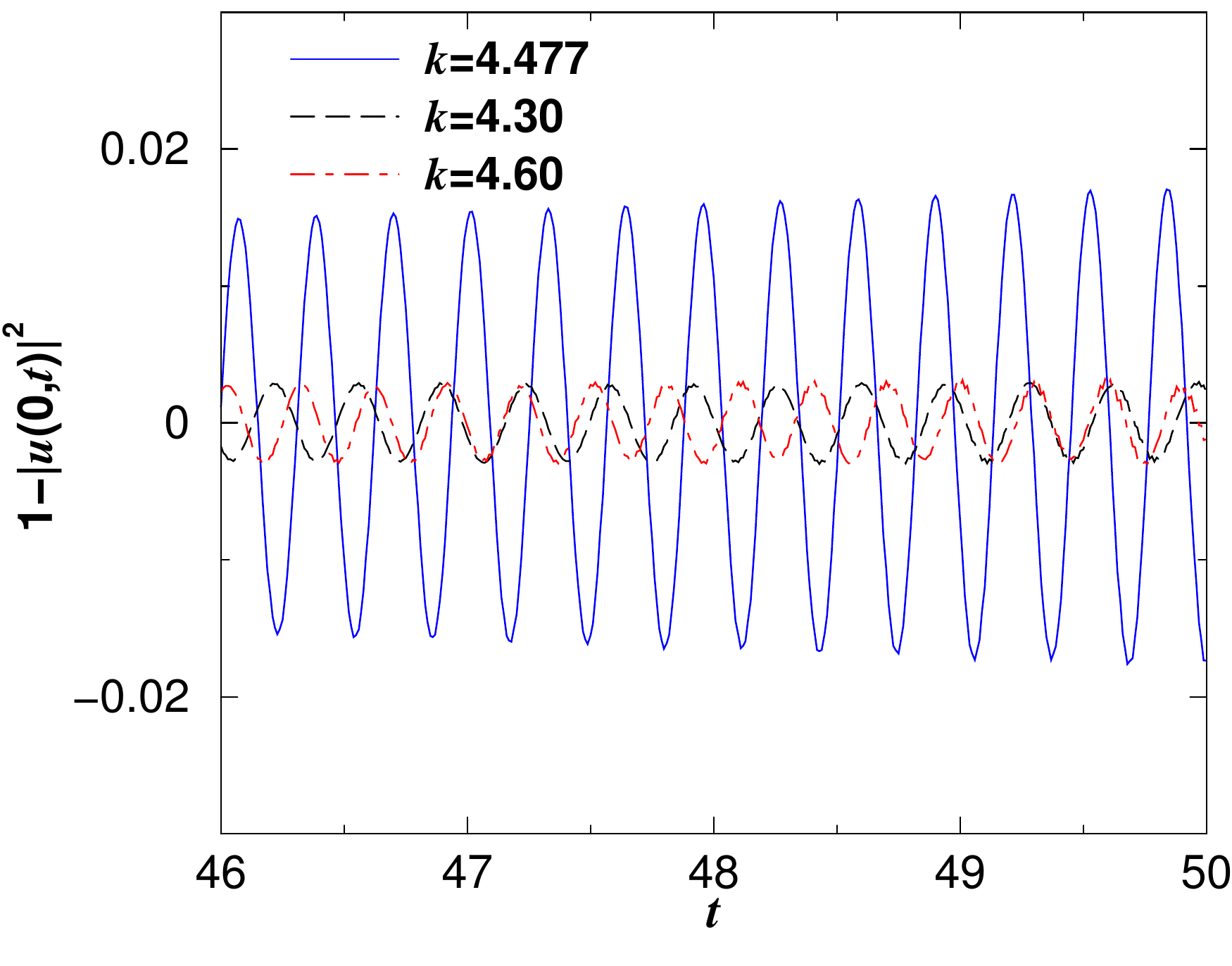}
}\caption{(color on-line) Following the results presented in Fig.\ref{f05}, here we
consider the second parametric resonance ($\omega=40$), from full-numerical 
calculations. As before, we follow the analytical prediction for the 
resonance, which is close to $k=4.477$. We note, in this case, that the 
rate of increasing of the amplitude at the resonance is not so fast as in 
the case of the first resonance. In the right panel we present two 
different small time intervals, where we are comparing the results 
obtained at the resonance with other two values of $k$. 
As before, the other parameters are in dimensionless units and such that 
$\gamma_0=0$, $\gamma_1=0.5$, $\eps_0=0.001$, $A=1$, and $c=1$.  
}
\label{f06}
\end{figure}
\begin{figure}[tbph]
\centerline{\hspace{0.5cm}
\includegraphics[width=10cm,clip]{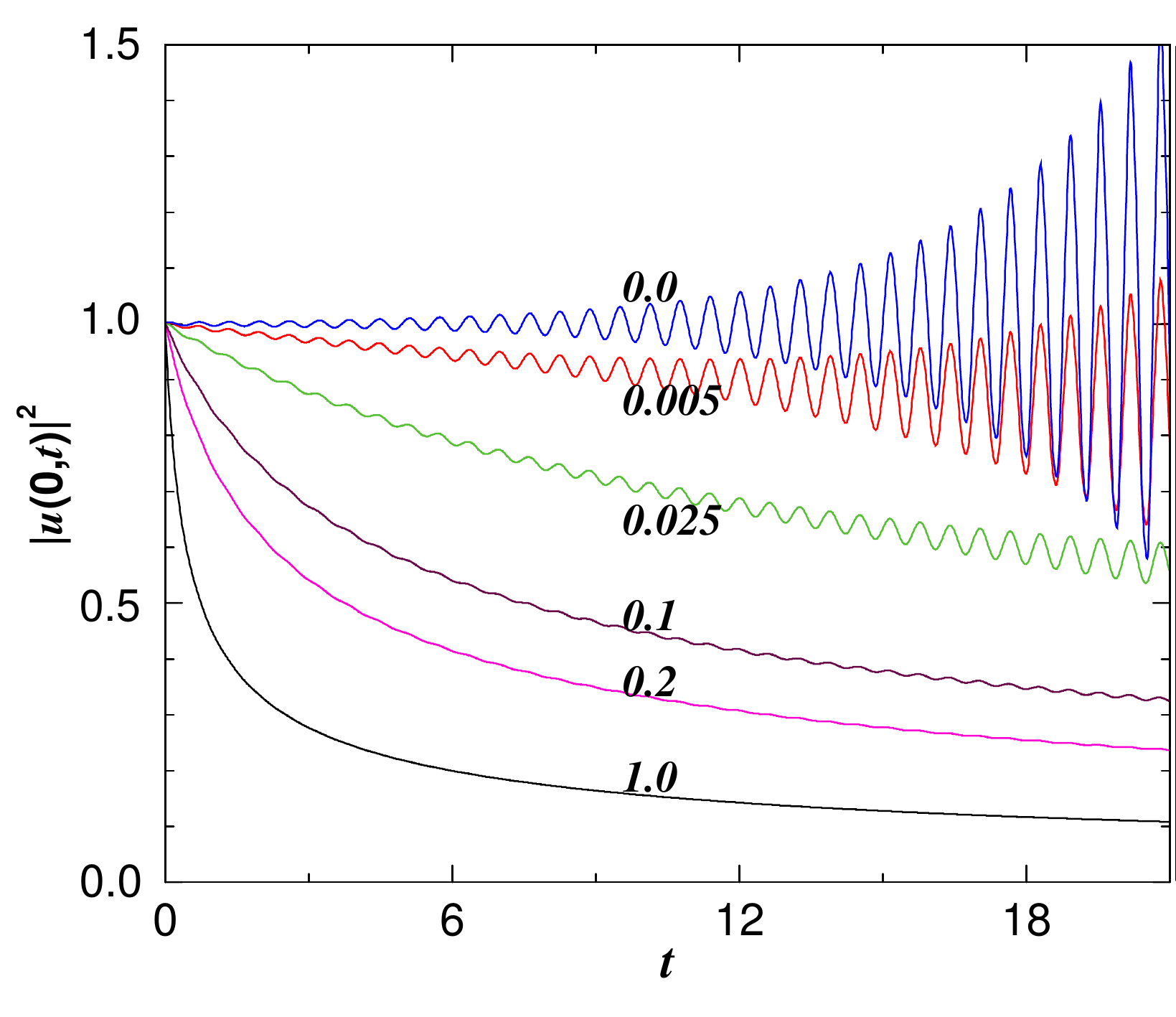}
}\caption{(color on-line) The effect of dissipation in the system we can exemplify with the results
presented in Fig.~\ref{f05}, by considering $k=3.17$ at the resonant position.
For that, in our full numerical calculation, we add in the quintic parameter $g$ a dissipative
imaginary term $\kappa_3$, varying it from zero (non-dissipative case shown by
the upper results) to $\kappa_3=1$ (lower curve), as indicated inside the frame. 
As expected, the amplitude of the 
resonance decreases gradually as we increase the dissipation.
The other parameters are the same as given in Fig.~\ref{f05}, in dimensionless units.
}
\label{f07}
\end{figure}
\begin{figure}[tbph]
\centerline{\hspace{0.5cm}
\includegraphics[width=7cm,clip]{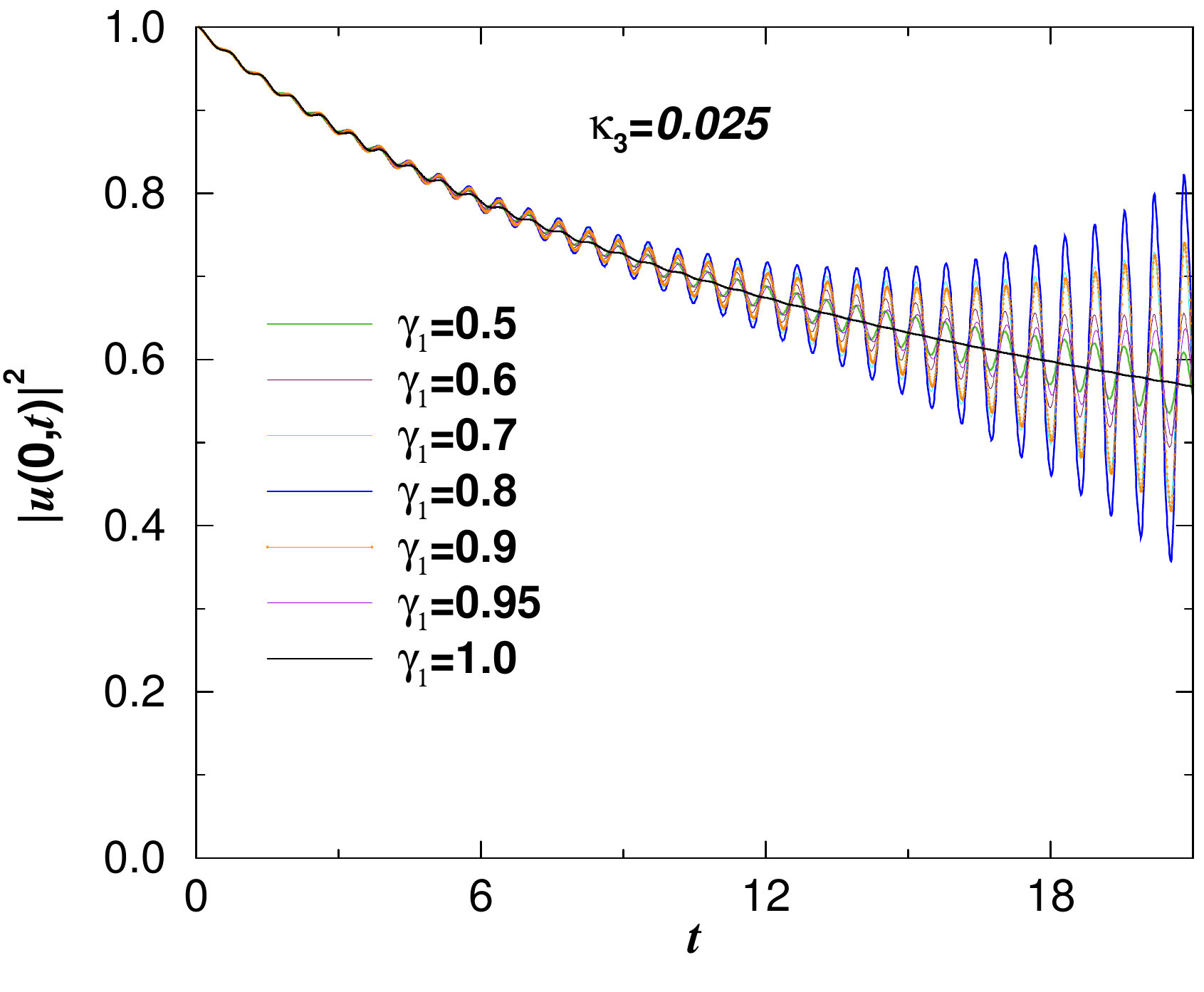}
\includegraphics[width=7cm,clip]{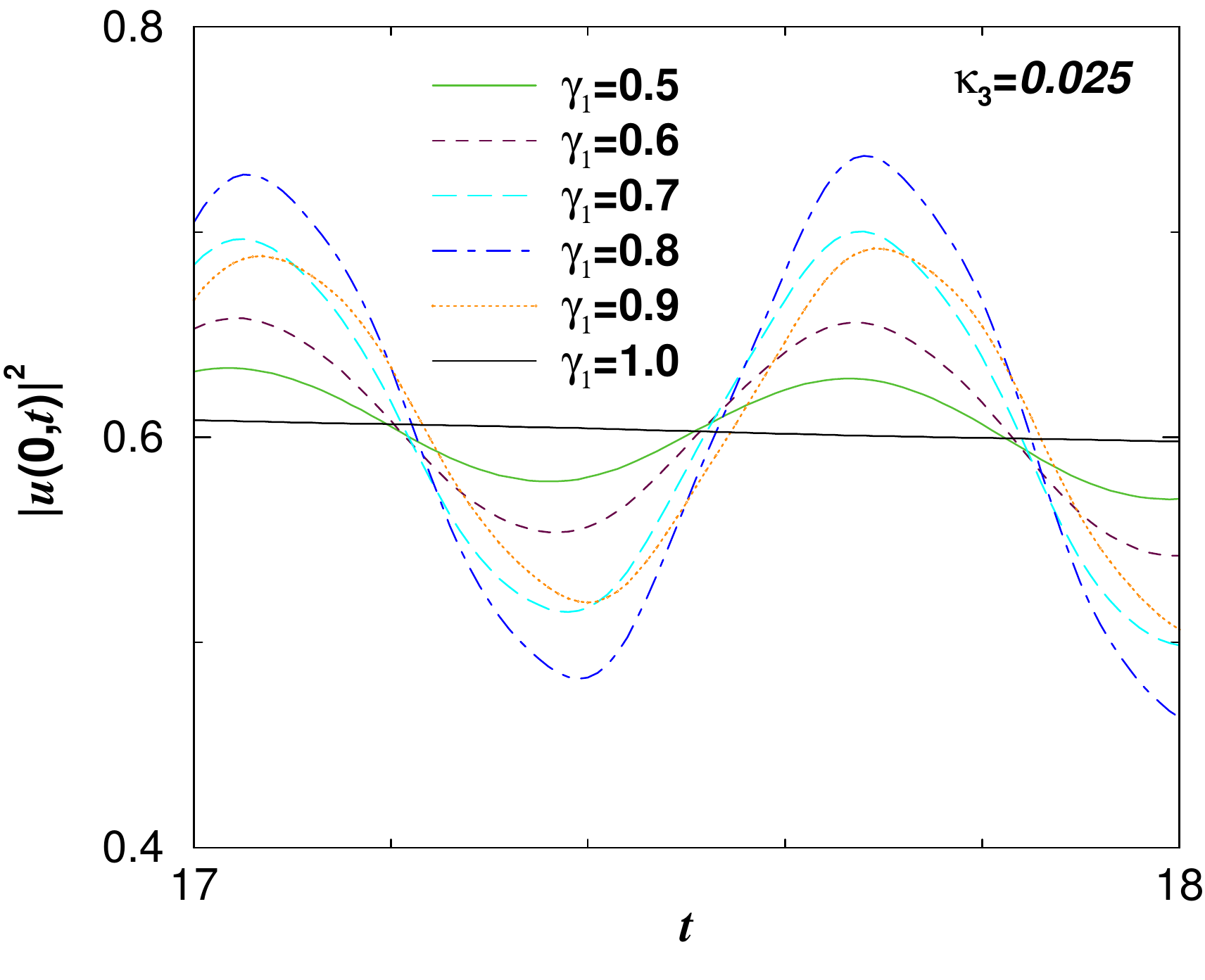}
}\caption{(color on-line) In this two panels we show that the effect of dissipation in the system can
be compensated by varying the parameter $\gamma_1$. For that, we follow
Fig.~\ref{f07}, for $k=3.17$, selecting the case where the dissipation parameter
is $\kappa_3=0.025$. The parameter $\gamma_1$ was varied, as shown inside
the frame, from 0.5 (same value as in Fig.~\ref{f07}), to 1. We noticed
that the maximum occurs near $\gamma_1 = 0.8$. The panel in the right, for a
small time interval, is for an easy identification of the different curves.
The other parameters are the same as shown in Fig.~\ref{f05}.
}
\label{f08}
\end{figure}
\begin{figure}[tbph]
\centerline{\hspace{0.5cm}
\includegraphics[width=7cm,clip]{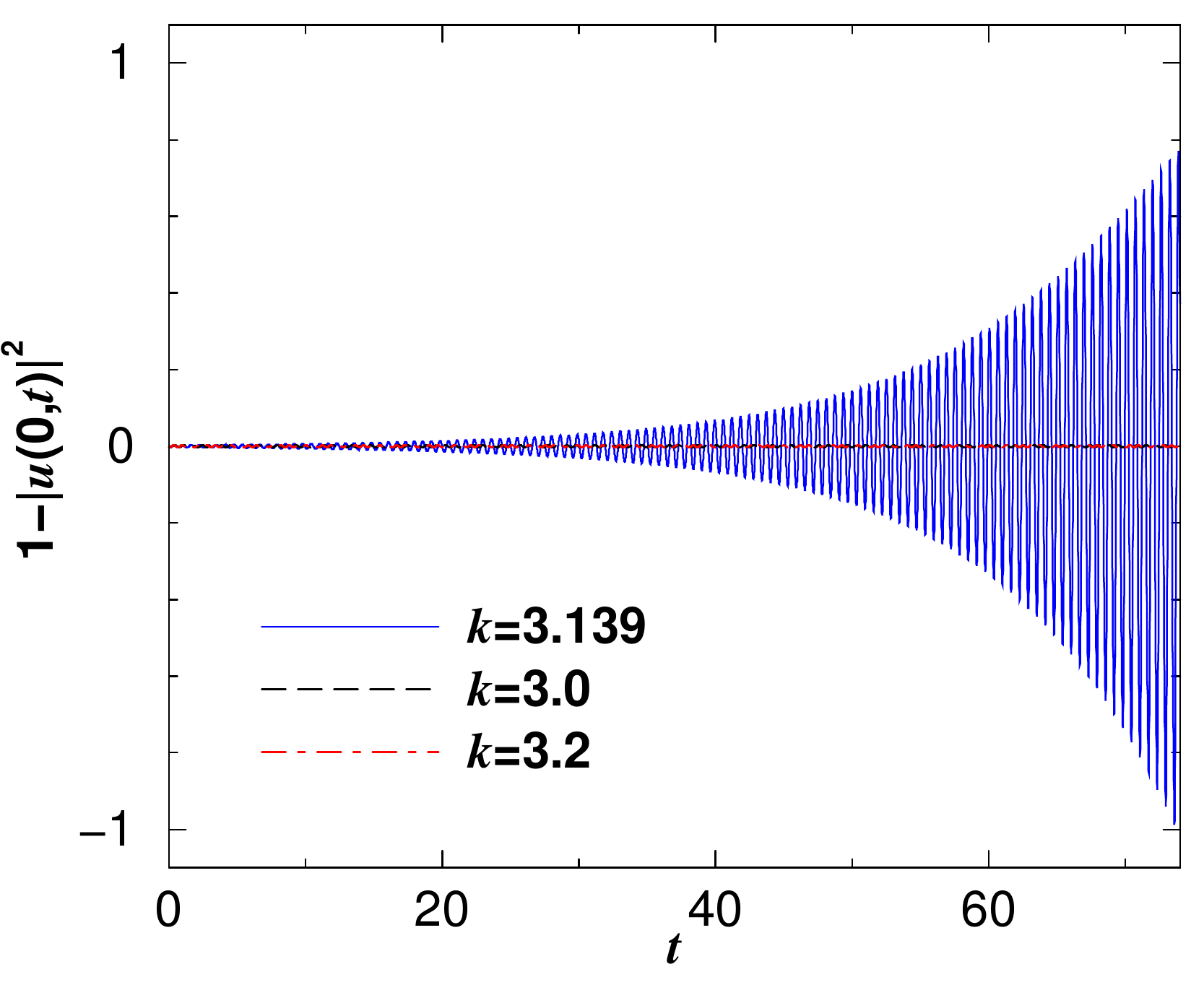}
\includegraphics[width=7cm,clip]{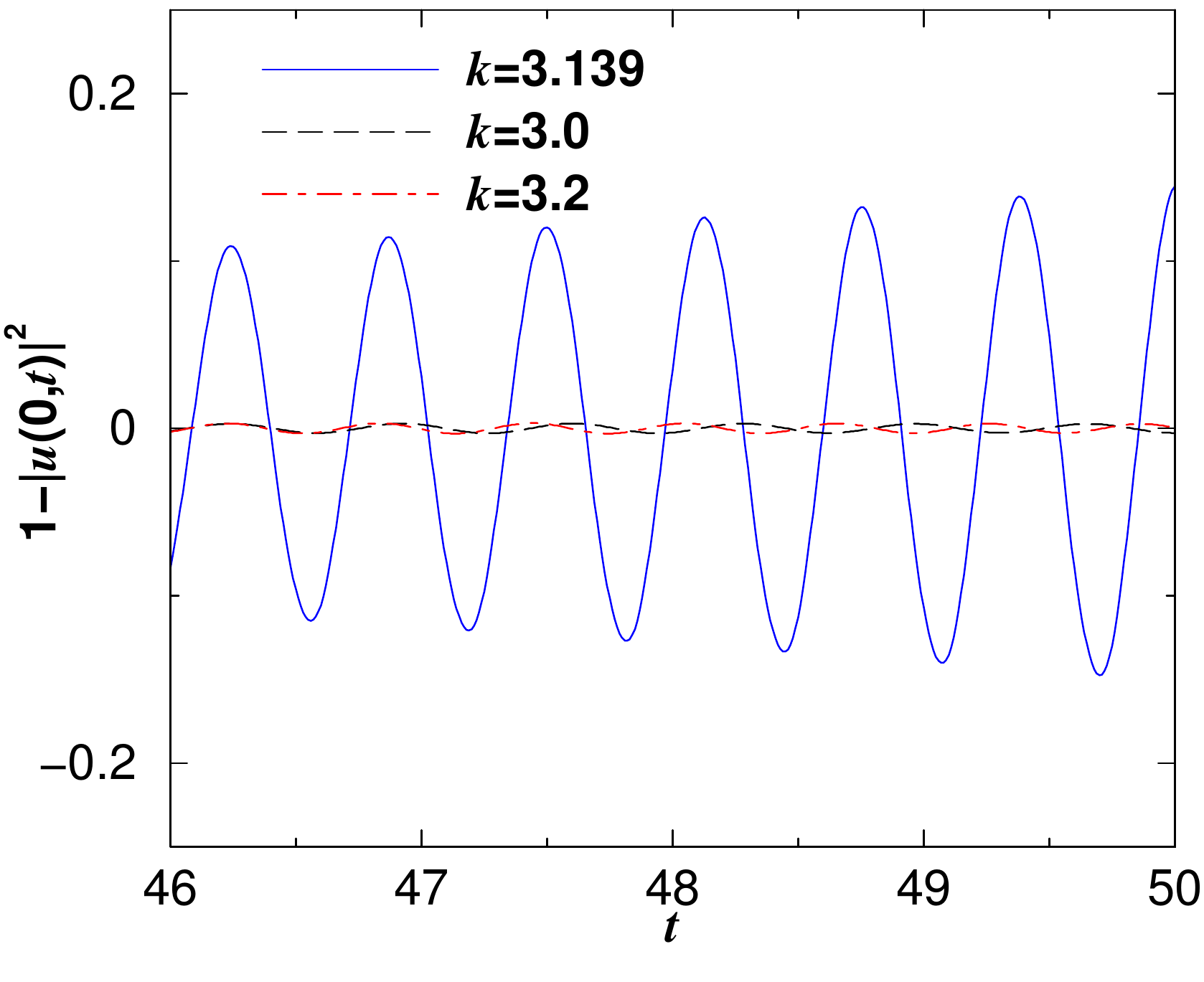}
}\caption{(color on-line) 
For the repulsive case, with $\gamma_0=-0.2$ and with $\gamma_1=0.2$, from full
numerical results, we present the case when the three-body interaction $g(t)$ is
given by Eq.(\ref{eq29}) (quartic case). The first parametric resonance 
for $\omega=20$, $\eps_0=0.001$, $A=1$, and $c_E=+1$ (such that $g(t)>0$), is found 
at $k=3.139$ (in agreement with prediction). 
In the right panel, for a small interval of time, we show how the amplitude
is changing for a small variation of the parameter $k$.
All quantities are dimensionless.
}
\label{f09}
\end{figure}

\begin{figure}[tbph]
\centerline{\hspace{0.5cm}
\includegraphics[width=7cm,clip]{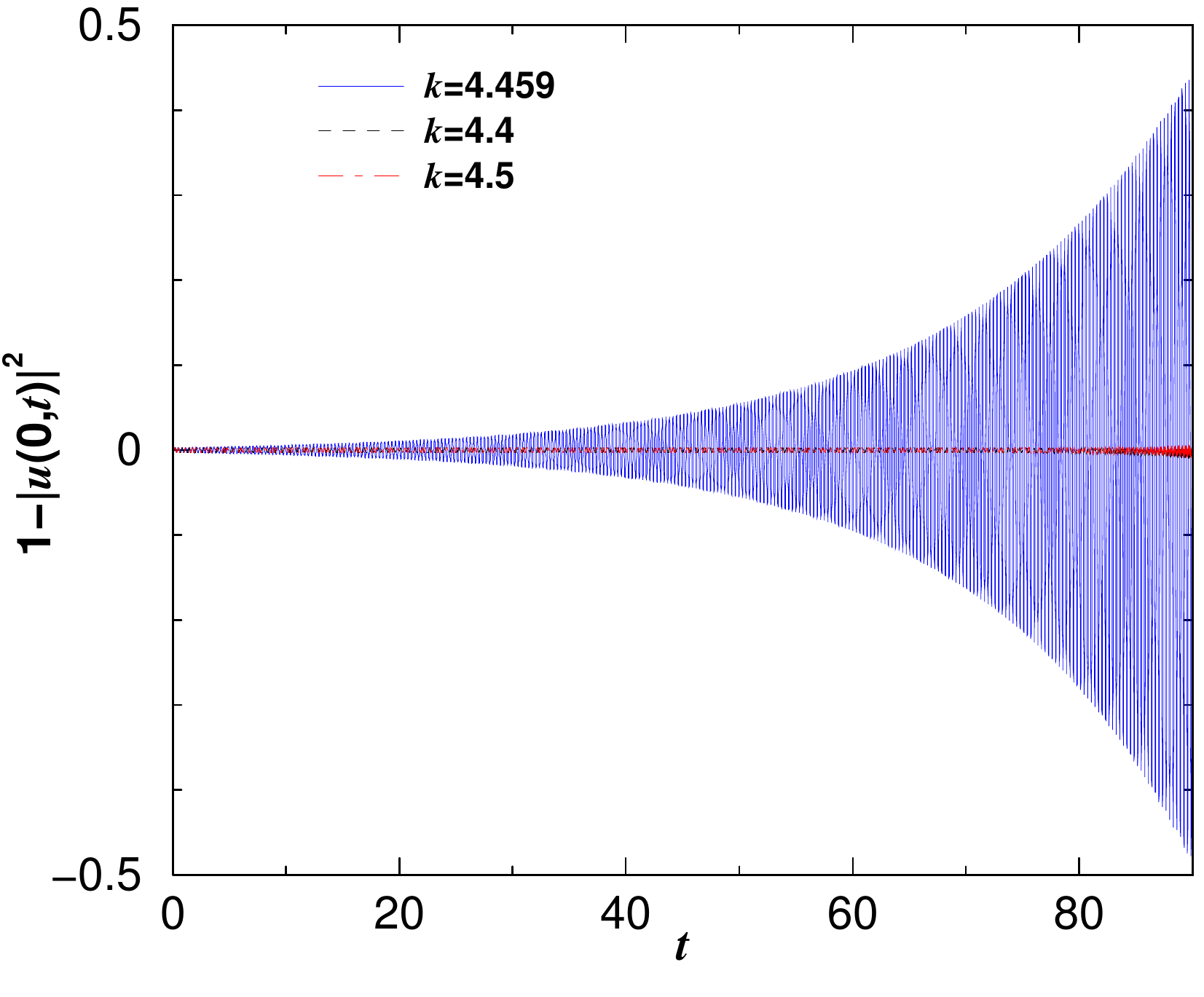}
\includegraphics[width=7cm,clip]{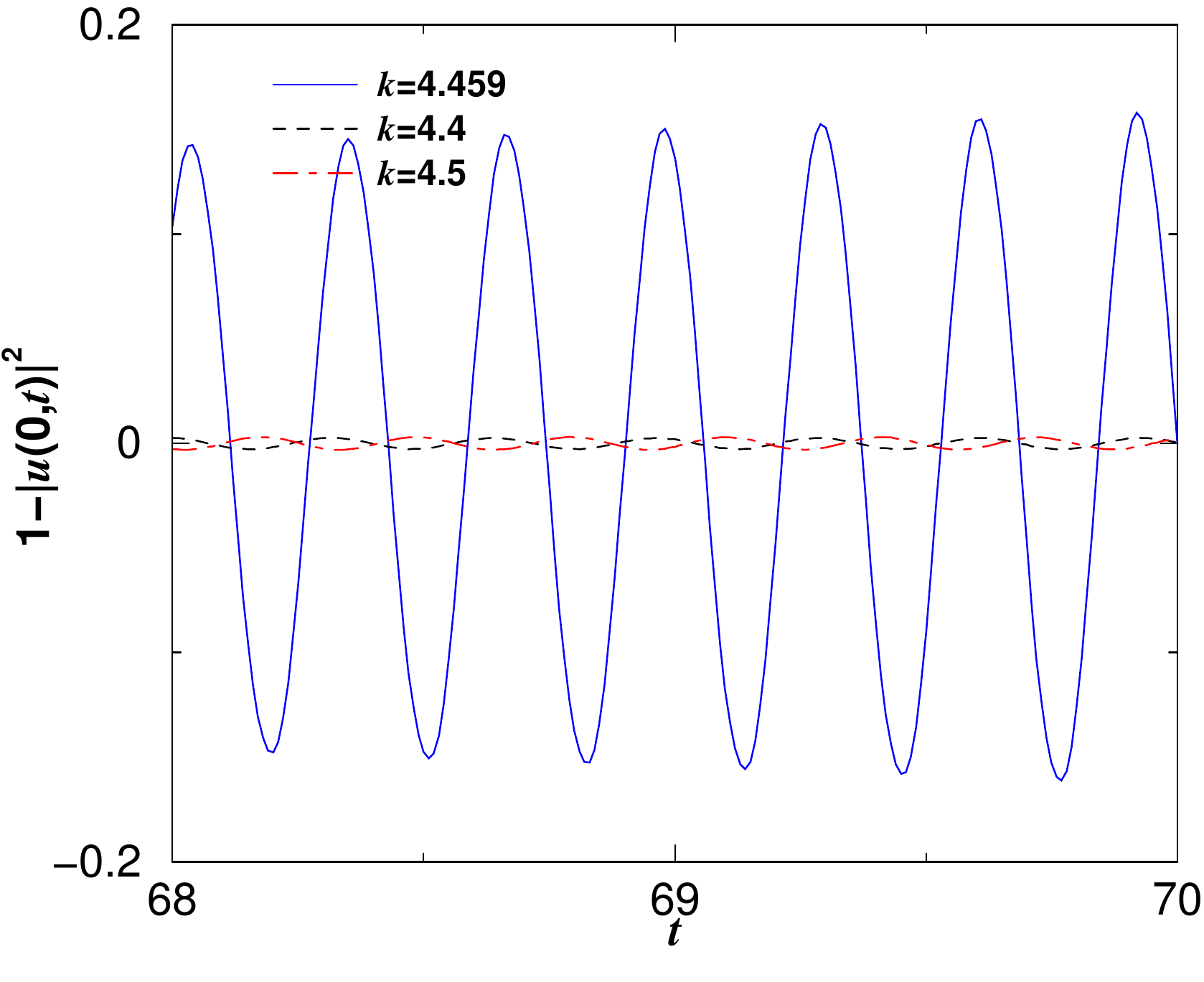}
}\caption{(color on-line) Following  Fig.~\ref{f09}, we show the corresponding second 
parametric resonance ($\omega=40$) for the repulsive two-body interaction ($\gamma_0=-0.2$), 
when we have positive three-body parameter ($c_E=+1$), in the quartic case. The resonance, 
as predicted, appears at $k=4.45$. This is shown by comparing with results obtained for $k$ smaller and larger than
this value, when the oscillation patterns remain almost constant (see right panel). In this
case, as compared with Fig.~\ref{f09}, the peak of the resonant value is manifested for larger 
values of $t$. All quantities are dimensionless and, except for $k$ and $\omega$,
the other parameters are given in  Fig.~\ref{f09}. 
}
\label{f10}
\end{figure}

\begin{figure}[tbph]
\centerline{\hspace{0.5cm}
\includegraphics[width=7cm,clip]{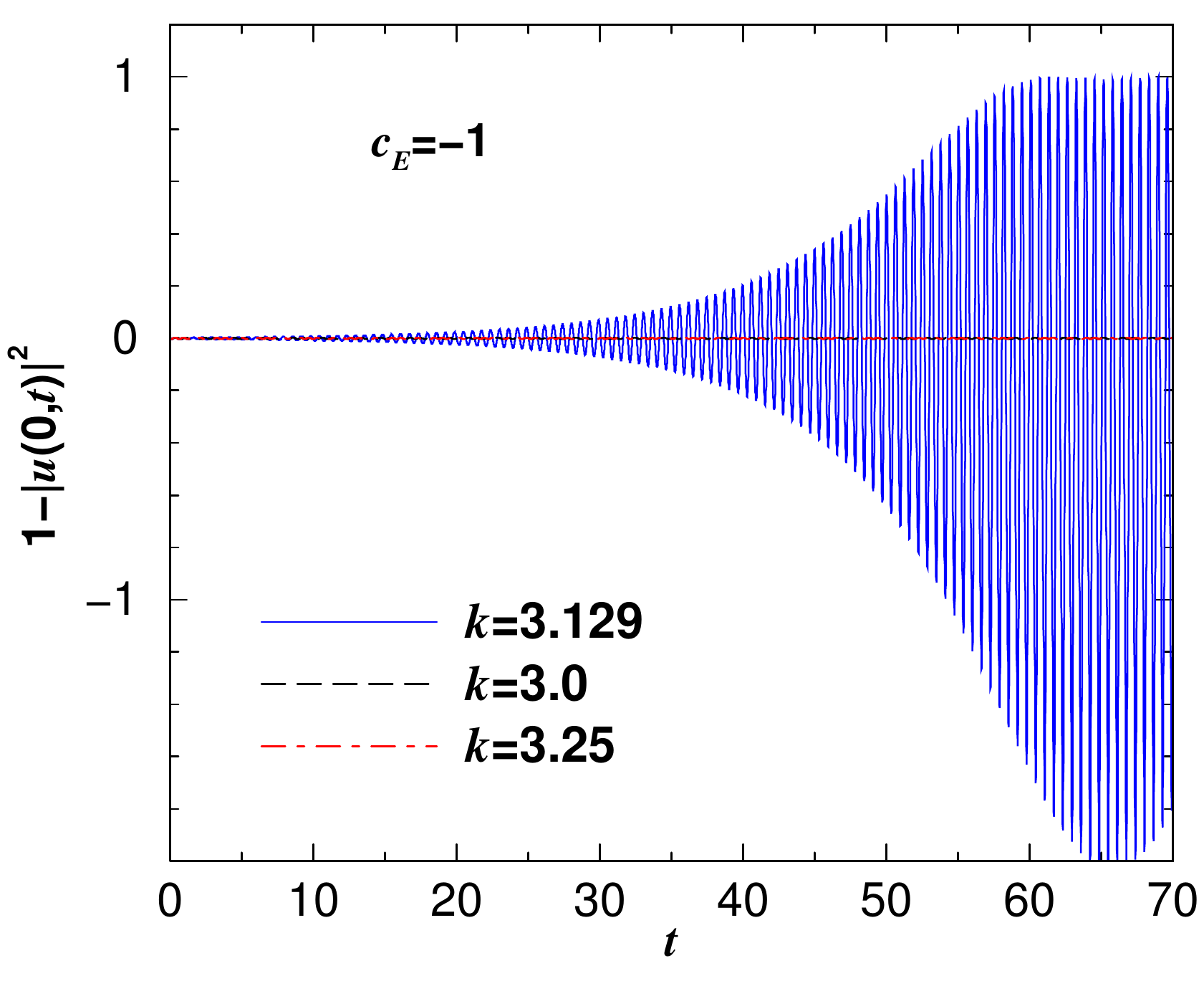}
\includegraphics[width=7cm,clip]{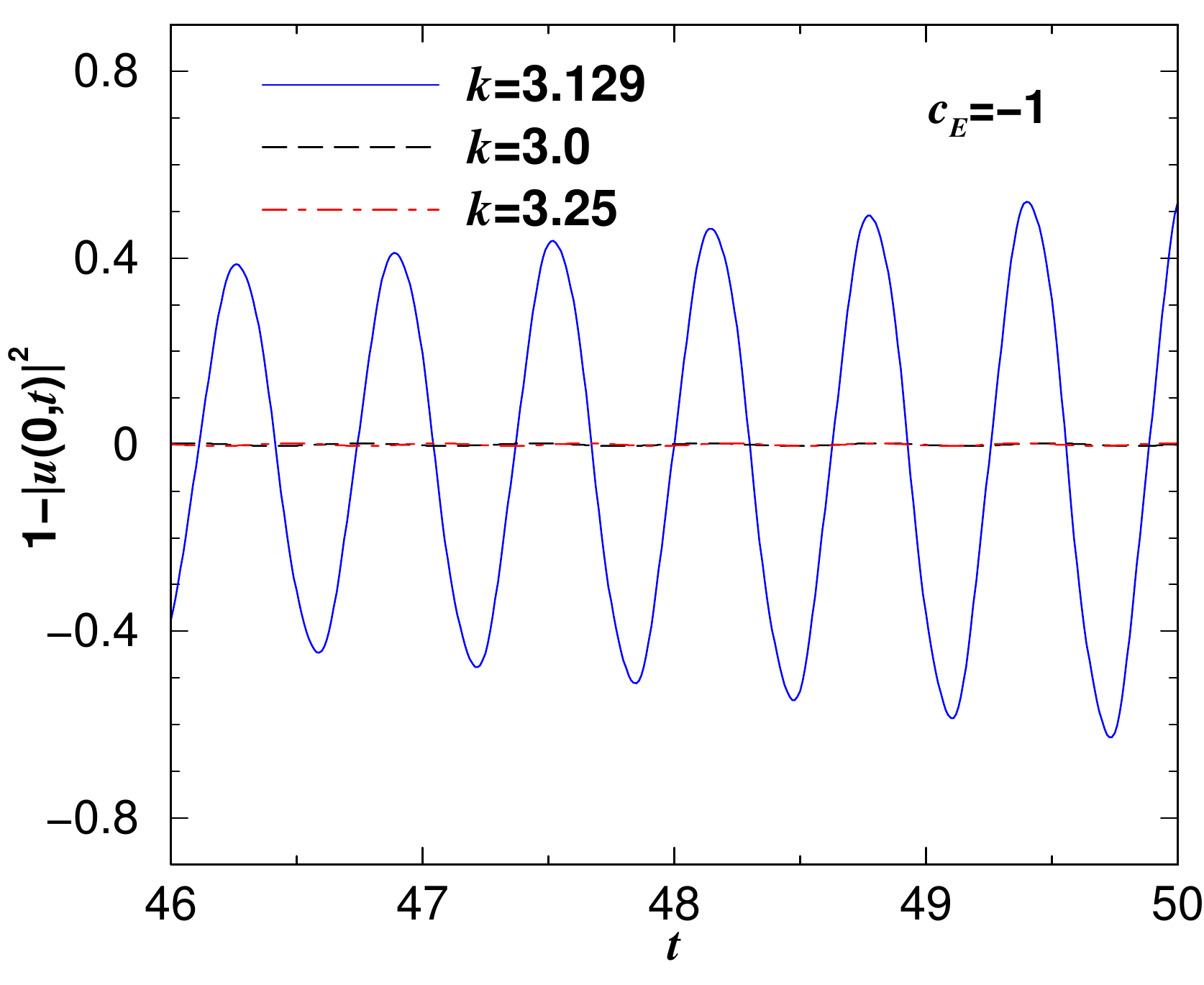}
}\caption{(color on-line) This figure presents the first parametric resonance, with $\omega=20$,
for the quartic case [see Eq.~(\ref{eq29})], with repulsive two-body interaction 
($\gamma_0=-0.2$), following the same dimensionless parametrisation as in Fig.~\ref{f09}, 
except that the sign of the three-body parameter $g(t)$ is inverted with $c_E=-1$. The resonance, 
as predicted, appears at $k=3.129$, with the right panel showing that density oscillation
remains almost constant for smaller and larger values of $k$. 
By comparing with Fig.~\ref{f09}, the resonance is manifested at smaller 
values of $t$ in case of $g(t)<0$. 
}
\label{f11}
\end{figure}

\section{Conclusion}
\label{sec:conclude}
 In this work we have investigated the generation of Faraday pattern in a BEC system, by engineering the time dependent 
three-body interactions. Two models were analysed, according to the mechanism of modulation and behaviour of the three-body 
interaction with respect to the atomic scattering length $a_s$. First, we have considered the strength of the three-body interaction as related to the square of $a_s$, supported by the model of Ref.~\cite{Tiesinga}.
Next, we study the generation of FW in the condensate when the strength of the three-body interaction is proportional to the fourth power of the atomic scattering length, which is valid for large values of $a_s $, near the Efimov 
limit~\cite{Bulgac,Braaten,Braaten2003}. 

The results of our analysis and numerical simulations show that the time-dependent three-body interaction excites 
Faraday patterns with the wavenumbers defined not only by $a_s$ and modulation frequency, but also by the 
amplitude of such oscillation. In the case of rapidly oscillating interactions, we derive the averaged GP equation 
by considering effective attractive three-body interactions. The MI analysis showed that the attractive three-body 
interaction effects are weakened by the induced modulations of nonlinear quantum pressure. 
In our analysis we have considered  both cases of repulsive and attractive two-body interactions.  
We also present simulations for repulsive three-body interactions in the quartic case, when it is proportional 
to the fourth power of $a_s$, considering the case of repulsive two-body interaction, where the behaviour of the
resonances can be well identified in agreement with predictions.
In all the cases the resonance positions can be easily verified with the help of analytical expressions.

For the experimental observation of Faraday waves it is important the case of the repulsive two-body interactions, since 
in the attractive case the initial noise, which can be originated from thermal fluctuations, can initiate the modulational 
instability, competing to the parametric one. Analytical predictions derived in the present work are in good agreement with 
results of numerical simulations, considering full time-dependent cubic-quintic extended GP equation. 

\section*{Acknowledgments}
\label{sec:ack}
F.A. acknowledges the support from Grant No. EDW B14-096-0981 provided by IIUM(Malaysia) and 
from a senior visitor fellowship from CNPq. AG and LT also thank the Brazilian agencies FAPESP, CNPq and CAPES for 
partial support.\\

\end{document}